\documentclass[preprint,12pt,authoryear]{elsarticle}

\usepackage{amssymb}

\usepackage{tikz}
\usepgflibrary{arrows}
\usepackage{subfigure}

\journal{Journal of Theoretical Biology}

\begin{document}

\newcommand{\E}{{\rm \bf E}}
\newcommand{\Var}{{ \rm \bf Var}}
\newcommand{\Cov}{{\rm  \bf Cov}}
\newcommand{\Eq}{\begin{equation}}
\newcommand{\Eeq}{\end{equation}}
\newcommand{\Eqn}{\begin{eqnarray}}
\newcommand{\Eeqn}{\end{eqnarray}}

\begin{frontmatter}

\title{Stochastic noise reduction upon complexification: positively correlated birth-death type systems}

\author[label 1]{Marianne Rooman}
\author[label 1]{Jaroslav Albert}
\author[label 2]{Mitia Duerinckx}

\address[label 1]{BioModeling, BioInformatics \& BioProcesses, Universit\'e  Libre de
Bruxelles, avenue Roosevelt 50, CP165/61, 1050 Brussels, Belgium}
\address[label 2]{Department of Mathematics, Universit\'e  Libre de
Bruxelles, boulevard du Triomphe, 1050 Brussels, Belgium}

\begin{abstract}
Cell systems consist of a huge number of various molecules that display specific patterns of interactions, which have a determining influence on the cell's functioning. In general, such complexity is seen to increase with the complexity of the organism, with a concomitant increase of the accuracy and specificity of the cellular processes. The question thus arises how the complexification of systems -- modeled here by simple interacting birth-death type processes -- can lead to a reduction of the noise -- described by the variance of the number of molecules. To gain understanding of this issue, we investigated the difference between a single system containing molecules that are produced and degraded, and the same system -- with the same average number of molecules -- connected to a buffer. We modeled these systems  using  It\=o stochastic differential equations in discrete time, as they allow straightforward analytical developments. In general, when the molecules in the system and the buffer are positively correlated, the variance on the number of molecules in the system is found to decrease compared to the equivalent system without a buffer. Only buffers that are too noisy by themselves tend to increase the noise in the main system. We tested this result on two model cases, in which the system and the buffer contain proteins in their active and inactive state, or protein monomers and homodimers. We found that in the second test case, where the interconversion terms are non-linear in the number of molecules, the noise reduction is much more pronounced; it reaches up to 20\% reduction of the Fano factor with the parameter values tested in numerical simulations on an unperturbed  birth-death model. We  extended our analysis to two arbitrary interconnected systems, and found that  the sum of the noise levels in the two systems generally decreases upon interconnection if the molecules they contain are positively correlated.
\end{abstract}


\end{frontmatter}


\section{Introduction}
\label{Introduction}

Biological systems involve large amounts of different  molecules that are closely packed in a relatively small area --- the cell and the intercellular medium. These molecules are located in some specific regions of space --- inside or outside the cell, inside or outside the nucleus, etc --- or move from one region to another. They interact in a specific manner to form transient or permanent complexes that perform the biological activity. These highly complex systems are moreover very sensitive to the environment (presence of other molecules) and external conditions (temperature, pH, salt concentration, etc). It is obviously impossible to take all these degrees of freedom into account. Therefore deterministic models can only reproduce the average of variables involved in biological processes. To gain insight into the actual time evolution of an individual process, stochastic models must be used, such as stochastic differential equations (SDE) or the master equation formalism. 

In spite of their highly complex and stochastic behavior, biological systems work very precisely and efficiently and perform their activity quite specifically, with a surprisingly low level of error. A striking observation is that while the overall complexity of the  cellular processes (for example the transcription machinery)  tends to increase with the complexity of the organisms (for example prokaryotes versus higher eukaryotes), the specificity and accuracy of these processes appear in general to increase too. In other words, the noise at the molecular and cellular levels tends to decrease when the number of degrees of freedom and thus the complexity of the organism increases. 

Note however that this overall tendency is not always true: some noise is  not  detrimental to biological systems. Sometimes it can create the diversity needed for cellular adaptation to, for example, different environments thereby generating new gene expression patterns or phenotypes \citep{Samollov, Thattai}. Also, cell differentiation has been suggested to be noise-driven \citep{Hoffmann, Forde}.

Intrinsic noise reduction in biological systems has  been investigated earlier by combinations of analytical  and numerical approaches. In particular, in the framework of gene expression networks, it has been shown that negative feedback can dramatically reduce the variability in gene expression \citep{Gardner,Becskei,Paulsson,Yi}. Actually, negative translational feedback appears to have a much greater efficiency at reducing stochasticity than negative transcriptional feedback \citep{Swain}. Also, complex promotor architectures are suggested to make gene expression regulation more precise \citep{Muller}. In contrast, in a genetic switch model consisting of a single gene with positive autoregulation, larger numbers of activator sites appear to lead to less accurate delays \citep{Albert}; the effect of cooperative binding of activators has also been studied and the level of noise seems to increase with the interaction energy \citep{Gutierrez}. Furthermore, cell-cell communication appears to lead in some (but not all) cases to decreased noise, due to the summation of the effects of all cells of the population \citep{Tanouchi, Weber, Koseska}. Finally, at the protein level, noise control is achieved through oligomerization \citep{Ghim, Bundschuh} or through the interaction between proteins and background molecules \citep{Morishita}.

To gain understanding about these issues, which are central for elucidating the basis of biological evolution but also for engineering novel cells in the framework of synthetic biology, we studied analytically a simple system containing molecules that are produced and degraded and compared it with the slightly more complex system in which the original system is connected to a second system --- called buffer. The system-buffer pair may be viewed as representing molecules that go from one region to the other, for example, from the cytoplasm to the nucleus and back. Also, molecules in the main system can be considered as being in their inactive state and those in the buffer in their active state due to their binding to a ligand. Alternatively, the molecules in the main buffer can be protein monomers and those in the buffer homomultimers. 

Our goal here is to compare the variance of the number of molecules -- that represents the noise -- of  a system with and without a buffer. We would like to emphasize that this comparison is performed for an equal average number of molecules in the main system (excluding the buffer). We indeed assume  that a biological system needs a fixed mean number of molecules to function correctly, whether or not a buffer is present.

We modeled the systems using discrete-time stochastic differential equations (SDE), in which the stochasticity is reproduced through Wiener processes. This formalism has the advantage of allowing easy analytical developments, which allow gaining basic understanding of the reasons underlying the noise reduction upon increase of complexity. For the sake of completeness, the link between this type of formalism and the Fokker-Planck equation and with the master equation is recalled explicitly. This clarifies the significance of the parameters that enter in the two approaches.    

\section{Stochastic system without a buffer}
Consider first a simple biological system consisting of molecules of type $\tilde y$ which are produced at some rate $\tilde P$ and eliminated at some other rate $\tilde D$ (see Fig.\ref{fig1}(a)). These molecules may for example be viewed as proteins that enter the system after translation from RNA and  leave it due to degradation, transformation or interaction with other biomolecules. They may also be seen as proteins that enter and leave a given cell or cell compartment. As biological processes are inherently stochastic, the amount of molecules, denoted by $\tilde Y$, and their production and degradation rates are taken as stochastic processes, defined on some probability space and indexed by a parameter $t$ that represents the time and varies over the interval $[0, T]$. A natural model for the time evolution of such a system consists of an It\=o stochastic differential equation in continuous time of the following form (see for example \cite{Allen}):
\Eq
d \tilde Y(t)=d \tilde P(t) - d \tilde D(t)   , \label{A}
\Eeq
where we assume that the production and degradation rates are each expressed as the sum of a deterministic term with drift coefficient denoted by $p^{(m)}$ and $d^{(m)}$, respectively, and of a stochastic term with diffusion coefficient $\sqrt {p^{(v)}}$ and $\sqrt {d^{(v)}}$ (where the superscripts $m$ and $v$ stand for "mean" and "variance"):
\Eqn
d \tilde P(t) & = & \tilde p^{(m)} (t, \tilde Y)  \, dt + \sqrt {\tilde p^{(v)} (t, \tilde Y)}\, d\tilde \eta(t)    , \nonumber \\ 
d \tilde D(t) & = & \tilde d^{(m)} (t, \tilde Y) \, dt + \sqrt {\tilde d^{(v)} (t, \tilde Y)} \, d\tilde \chi (t)    ; \label{A1} 
\Eeqn
$ \tilde \eta(t)$ and $\tilde \chi(t)$ stand for two independent Wiener processes. Remember that, by definition, $\tilde\eta(0)=0$ and $\tilde\chi(0)=0$, and that both $\tilde \eta(t)-\tilde\eta(t')$ and $\tilde\chi(t)-\tilde\chi(t')$ follow a $\mathcal N(0,t-t')$ distribution for all $t,t'$. Note also that the Wiener process has continuous-valued realizations and is thus appropriate when $\tilde Y$ represents concentrations of molecules, or when the number of molecules is large enough to be approximated as a continuous variable, whereas Poisson processes would be better suited when  $\tilde Y$ represents small numbers of molecules. We will consider here for simplicity  only Wiener processes, with $\tilde Y$ taking positive real values and corresponding to large numbers of molecules \citep{Allen}.

In general, the drift and diffusion coefficients may depend on $\tilde Y$. It has been shown that the  Fokker-Plank equation for a production process corresponds to an It\=o SDE with drift and diffusion coefficients independent of $\tilde Y$, whereas for a degradation process the drift coefficient is proportional to $\tilde Y$ and the diffusion coefficient to $\tilde Y^{1/2}$ \citep{Allen}. Hence, we naturally set:
\Eqn 
d \tilde P(t) & = & \tilde p^{(m)} \, dt +  \sqrt {\tilde p^{(v)}} \, d\tilde \eta(t)      , \nonumber \\ 
d \tilde D(t)   &=&  \tilde d^{(m)} \, \tilde Y(t) \, dt + \sqrt { \tilde d^{(v)}  \, \tilde Y(t)}  \, d\tilde \chi(t)   . \label{A2}
\Eeqn
We assumed here that the production and degradation parameters are time independent. This makes the subsequent calculations simpler but is actually unnecessary; we only have to assume that these parameters allow for a long-time limit as $t\to\infty$. Note that we chose to model our  system using It\=o SDEs rather than Stratonovich SDEs  because of the similarities of the former with the Fokker-Planck  equation (see {\it e.g.} \cite{Allen, Allen1}), known to yield relevant descriptions of biological systems.

For the simplicity of the subsequent calculations, we approximate the  continuous SDE given by Eq.(\ref{A}) by a discrete-time SDE, where the time interval $[0,T]$ is subdivided in $N$ equal-length intervals $0=t_0< \ldots< t_N=T$, with $t_n=n \Delta t$ and $\Delta t = T/N$. Using Milstein's discretization method \citep{Milstein}, we get:
\Eq
\tilde Y_{n+1}=\tilde Y_n +  \Delta \tilde P_n -  \Delta \tilde D_n  
  + \Delta \tilde M_n   ,\label{B}
\Eeq
for all positive integers $n \in [0,N]$, where the discretized production and degradation rates, and the Milstein term $\Delta M_n$, are given by:
\Eqn 
 \Delta \tilde P_n  & = & \tilde p^{(m)} \Delta t + \sqrt {\tilde p^{(v)}  } \,  \Delta \tilde \eta_n   ,   \nonumber \\ 
 \Delta \tilde D_n   &=& \tilde d^{(m)}  \, \tilde Y_n \, \Delta t + \sqrt  {\tilde d^{(v)}\,   \tilde Y_n }  \, \Delta \tilde \chi_n    ,  \nonumber \\
\Delta \tilde M_n  & = & - \frac 14 \tilde d^{(v)} ( (\Delta \tilde \chi_n)^2 - \Delta t)   ,   \label{B1} 
\Eeqn 
with $\tilde\eta_n=\tilde\eta(t_n)$ and $\Delta \tilde \eta_n= \tilde \eta_{n+1} - \tilde \eta_n$, so that in particular $\tilde \eta_0=0$, $\E (\Delta \tilde \eta_n)=0$ and $\Var (\Delta \tilde \eta_n)= \Delta t $, and similarly for $\tilde \chi$.  
Owing to the Milstein term $\Delta \tilde M_n$,  the mean square error between $\tilde Y_n$ and $\tilde Y(t_n)$ is of the order of $(\Delta t )^2$ (see {\it e.g.} \cite{Allen, Milstein}). The Milstein method is thus more accurate than the Euler-Maruyama method in which $\Delta \tilde M_n$ is set to $0$, and where the mean square error is of the order of $\Delta t$. Note  that $\E(\Delta \tilde M_n)=0$ and $\Var(\Delta \tilde M_n) =O (\Delta t)^2$, so that this correction term will not appear in our final results. It is, however, important for numerical simulations.

Computing the mean and variance of the discretized production and degradation rates (\ref{B1}) yields:
 \Eqn
 &\E(\Delta \tilde P_n)=\tilde p^{(m)} \Delta t   , & \qquad  \Var(\Delta \tilde P_n)=\tilde p^{(v)} \Delta t   ,   \nonumber \\ 
&\E(\Delta \tilde D_n)=\tilde d^{(m)} \, \E(\tilde Y_n) \, \Delta t   , & \qquad  \Var(\Delta \tilde D_n)=\tilde d^{(v)} \, \E(\tilde Y_n) \, \Delta t  ,  \label{B2}
  \Eeqn
up to the second order in $\Delta t$. It is now apparent that the superscripts $m$ and $v$ refer to the mean and the variance, respectively. The continuous-time equations (\ref{A}-\ref{A2}) are obtained by taking the limit $\Delta t \to 0 $, {\it i.e.} by taking the limit $N \to \infty$ while keeping $T$ constant. In what follows, we consider a fixed time-discretization level, which means that we keep $\Delta t$ small but constant. Furthermore, we assume the weak convergence of the system towards a steady state, $\tilde Y$, in the long-time limit, for small discretization step. More precisely, we assume that, for any sufficiently small fixed $\Delta t$, $\tilde Y_N$ converges weakly to some random variable $\tilde Y_{\Delta t}$ in the limit $T=N \Delta t \to\infty$, and that $\tilde Y_{\Delta t}$ then converges weakly to some random variable $\tilde Y$ in the limit $\Delta t\to0$. In what follows, when mentioning the steady state limit $X$ of a process $X_n$ when $n \to \infty$, we mean the so-defined limit (which we always assume to exist for the processes we consider --- a seemingly reasonable hypothesis in many cases of interest, excluding however systems having {\it e.g.} limit cycles or oscillatory behaviors). The weak convergence of a process $X_n \to X$ implies the convergence of the moments: $\E( X_n^p)\to\E (X^p)$ for any $p>0$, if  the $X_n$'s are bounded by some constant. Given that the processes $X_n$  considered here represent biomolecules in a biological (bounded) system, this is a reasonable assumption.

Let us now compute the mean and variance of the amount of molecules in the steady state limit. Taking the mean of Eq.(\ref {B}) yields:
\Eq
\E(\tilde Y)=\frac { \tilde p^{(m)}}{  \tilde d^{(m)}}  . \label{B2_1}
\Eeq
Taking the square of both members of Eq.(\ref{B}) gives the Fano factor in the steady-state limit:
\Eq
\frac {\Var (\tilde Y )}{\E (\tilde Y)}= \frac 12 \left ( \frac { \tilde d^{(v)} }{\tilde d^{(m)} } + \frac {\tilde p^{(v)} }{ \tilde p^{(m)}} \right )  . \label{B2_2}
\Eeq
By virtue of this relation, the equality of the mean and variance of the degradation rate and of the production rate ({\it i.e.}  $ \tilde d^{(v)} =\tilde d^{(m)}$ and $ \tilde p^{(v)} =\tilde p^{(m)}$) implies the equality of the mean and variance of the number of molecules in the steady state limit: $\Var(\tilde Y)=\E(\tilde Y)$; the Fano factor is thus equal to one. As seen in the next section, this result is actually in agreement with the finding that the master equation for a simple birth-death process yields a steady-state probability distribution of the number of molecules that is Poissonian \citep{Walczak}.

\section{Relation with the master equation formalism}
Before analyzing more complex systems using our discretized SDE models, it is informative to recall the link between this description and the often used master equation formalism. For that purpose, first consider the relation between It\=o SDEs, where the stochasticity is explicitly introduced through Wiener processes, and the Fokker-Planck equation, which is an equation for the probability density function (see {\it e.g.} \cite{Allen, Friedman, Gikhman}). In particular, the continuous-time version of the SDE considered here, given by Eqs(\ref{A}-\ref{A1}), is equivalent (under mild conditions) to the following Fokker-Planck equation:
\Eq
\frac {\partial P}{\partial t}=-\frac {\partial}{\partial \tilde y}{ \left ( (\tilde p^{(m)} - \tilde d^{(m)}\tilde y) P\right )}+ \frac12  \frac{\partial^2}{\partial \tilde y^2}{ \left ( (\tilde p^{(v)} + \tilde d^{(v)}\tilde y) P\right )}  ,\qquad t,\tilde y\ge0, \label{FP}
\Eeq
where $P \equiv P(t,\tilde y)$ is the probability density of the solution to the considered SDE. This equation resembles a diffusion equation, with an extra term that corresponds to a deterministic drift. Note that the parameters $p^{(v)}$ and $d^{(v)}$ of the Wiener processes in the SDE (\ref{A}) enter in the diffusion term of the Fokker-Planck equation (\ref{FP}), whereas the parameters $p^{(m)}$ and $d^{(m)}$ occur in the drift term.

The relation between the Fokker-Planck equation (where the number of particles is continuous) and the master equation (where it is discrete)  is well known and easy to obtain. In particular, the Fokker-Planck equation (\ref{FP}) corresponds to the following master equation:
\Eq
\frac {\partial P_{\tilde y}}{\partial t}=g_{\tilde y-1}P_{\tilde y-1} + r_{\tilde y+1}P_{\tilde y+1}-(g_{\tilde y}+r_{\tilde y})P_{\tilde y} , \qquad t\ge0,~\tilde y\in\mathbb N_0, \label{ME}
\Eeq
where $P_{\tilde y}(t)$ is the probability distribution obtained from $ P(t,\tilde y)$ through a discretization of the values $\tilde y$ of the number of molecules. Note that  this formalism is valid even for small number of particles, whereas the Fokker-Planck equation is a good approximation only when the number of particles is large enough. The production rate $g_{\tilde y}$ and the degradation rate $r_{\tilde y}$ are given in terms of the parameters of the original SDE (\ref{A}) as:
 \Eqn
 g_{\tilde y}&=&\frac 12 \left (\tilde p^{(v)}+\tilde p^{(m)}\right)+\frac 12 \left ((\tilde d^{(v)}-\tilde d^{(m)})\tilde y \right ), \nonumber \\
 r_{\tilde y}&=&\frac 12 \left ( (\tilde d^{(v)}+\tilde d^{(m)})\tilde y \right )+\frac 12 \left (\tilde p^{(v)}-\tilde p^{(m)}\right )
  . \label{ME1}
 \Eeqn
In a simple birth-death process, we have $g_{\tilde y}=g$ and $r_{\tilde y}=r \tilde y$, which amounts to  setting $\tilde p^{(v)}=\tilde p^{(m)}=g$ and $\tilde d^{(v)}=\tilde d^{(m)}=r$. The number of molecules $\tilde Y$ is in this case known to follow a Poisson distribution \citep{Walczak}. In particular, this implies $\Var(\tilde Y)=\E(\tilde Y)$, which is in agreement with the result (\ref{B2_2}) obtained with discretized SDEs.

When $\tilde p^{(v)}\neq \tilde p^{(m)}$ and/or $\tilde d^{(v)}\neq \tilde d^{(m)}$, the production and/or degradation rates contain two terms, one $\tilde y$-dependent and the other $\tilde y$-independent. The SDE of Eq.(\ref{A}) does not describe a simple birth-death process in this case. The extra terms can be interpreted as representing an external perturbation that the system undergoes, due {\it e.g.} to interactions with other molecules in the neighborhood. 

In the case $\tilde p^{(v)}>\tilde p^{(m)}$ and $\tilde d^{(v)}>\tilde d^{(m)}$,  the coefficients of the Wiener processes in the SDE (\ref{A}) are increased compared to the simple, unperturbed, birth-death process and thus  the noise level is increased. This larger noise is also reflected in the inequality $\Var (\tilde Y)> \E(\tilde Y)$, which follows from Eq.(\ref{B2_2}). In contrast, when $\tilde p^{(v)}<\tilde p^{(m)}$ and $\tilde d^{(v)}<\tilde d^{(m)}$, the coefficients of the Wiener processes are decreased and thus the noise level is reduced. We have in this case the inequality $\Var (\tilde Y)< \tilde Y$. 

At the limit of vanishing noise, {\it i.e.} when $\tilde p^{(v)}=0$ and $\tilde d^{(v)}=0$, we obtain $g_{\tilde y}=-r_{\tilde y}=\frac12(\tilde p^{(m)}-\tilde d^{(m)} \tilde y)$, so that the diffusion term in the  Fokker-Planck equation (\ref{FP}) vanishes, and so do the coefficients of the Wiener processes in the SDE (\ref{A}). The model is thus no longer stochastic but becomes purely deterministic, with $\Var (\tilde Y)=0$. The master equation  becomes simply a first-order differential equation: $\partial_t P_{\tilde y}=g_{\tilde y-1}P_{\tilde y-1} - g_{\tilde y+1}P_{\tilde y+1}$, which is solved by $P_{\tilde y}=\delta_{\tilde y,\E(\tilde Y)}$, implying $\tilde Y=\tilde p^{(m)}/\tilde d^{(m)}$ almost surely, under the condition that $\tilde p^{(m)}/\tilde d^{(m)}$ is an integer.

In summary, when considering a pure birth-death process, the parameters in the SDE (\ref{A}) (or in its discretization (\ref{B})) must satisfy the relations $\tilde p^{(v)}=\tilde p^{(m)}$ and $\tilde d^{(v)}=\tilde d^{(m)}$. Other parameter values describe systems undergoing external perturbations that do not occur in the most simple model. This may be the case for example when the molecules are produced or degraded through a process that involves other molecules, which are not taken into account explicitly in the model, but whose effect is encoded in effective production and degradation parameters. This point will be further discussed in the Conclusion section. In what follows, we do not impose any restrictions on the parameter values, and state that larger variances ({\it i.e.} $\tilde p^{(v)}>\tilde p^{(m)}$ and $\tilde d^{(v)}>\tilde d^{(m)}$) define more noisy production and degradation gates, while smaller variances ({\it i.e.} $\tilde p^{(v)}<\tilde p^{(m)}$ and $\tilde d^{(v)}<\tilde d^{(m)}$) define more deterministic gates.

\section{Stochastic system with a buffer}
Now consider the slightly more complex system illustrated in Fig.\ref{fig1}(b), consisting of two subsystems, containing molecules of type $y$ and $z$, respectively. These subsystems are connected, and molecules
of type $y$ convert into molecules of type $z$ and conversely; $y$ may for example be protein monomers
and $z$ protein multimers, or $y$ may be an inactive and $z$ an active state of the same protein, or $y$ may be located in one cell compartment and $z$ in another one. Molecules $y$ are produced and degraded, while  molecules $z$ are degraded but not produced (except through conversion from $y$); this is a realistic assumption with respect to the examples cited above. Such a system can naturally be modeled by the following coupled discretized SDEs:
 \Eqn 
 Y_{n+1} &=& Y_n +  \Delta  P_n -  \Delta D_n  - \alpha \, [\Delta F_n - \Delta G_n ] +\Delta M_n  ,\nonumber \\
 Z_{n+1} &=& Z_n    -  \Delta E_n + [ \Delta F_n - \Delta G_n ] + \Delta N_n  , \label{BC}
\Eeqn
where $Y$ and $Z$ stand for the number of molecules of type $y$ and $z$, respectively. The constant $\alpha$ represents the number of molecules of type $y$ that make up one molecule of type $z$; in particular,  $\alpha =2$ when $z$ are protein dimers.  The production term $\Delta P_n$, the degradation terms $\Delta D_n$ and $\Delta E_n$, and the interconversion terms $\Delta F_n$ and $\Delta G_n$ which convert molecules of type $y$ into molecules of type $z$ and conversely --- all generically represented by $\Delta B_n$ hereafter --- decompose into a deterministic drift part (proportional to $\Delta t$) and a  stochastic part (of the order of $\Delta t^{1/2}$):
\Eq 
 \Delta  B_n    =  b^{(m)} \, U(Y_n, Z_n)\, \Delta t + \sqrt {b^{(v)}  V(Y_n, Z_n) }\,  \Delta \beta_n   . \label{BC1}
\Eeq 
We suppose here for simplicity that $U(Y_n, Z_n)$ and $V(Y_n, Z_n)$ do not depend explicitly on the time. The discretized Wiener processes $\beta_n$ that describe the stochasticity of the  different  rates are all chosen independent of each other. As before, $\Delta \beta_n=  \beta_{n+1} -  \beta_n$, $ \beta_0=0$, $\E ( \Delta \beta_n)=0$, and $\Var (\Delta  \beta_n)= \Delta t $. The mean and variance of $\Delta B_n$ can be expressed as:
\Eqn 
 \E(\Delta  B_n) &=& b^{(m)} \,  \E(U(Y_n, Z_n)) \, \Delta t  , \nonumber \\ 
 \Var(\Delta  B_n) &=& b^{(v)} \, \E(V(Y_n, Z_n)) \, \Delta t  , \label{BC2}
\Eeqn
up to the second order in $\Delta t$. Similarly as in the case without a buffer, we have $U(Y_n, Z_n)=1=V(Y_n, Z_n)$ for the production rate ($\Delta P_n$),  $U(Y_n, Z_n)=Y_n=V(Y_n, Z_n)$ for the degradation rate ($\Delta D_n$) of molecules of type $y$, and $U(Y_n, Z_n)=Z_n=V(Y_n, Z_n)$ for the degradation rate ($\Delta E_n$) of molecules  $z$. For the interconversion rates ($\Delta F_n$ and $\Delta G_n$), $U(Y_n, Z_n)$ and $V(Y_n, Z_n)$ are, for the moment, left unspecified: they depend on the specific model considered. Writing generically $\Delta B_n=\{b^{(m)},b^{(v)},U(Y_n, Z_n),V(Y_n, Z_n),\beta_n\}$, we have thus $\Delta P_n=\{p^{(m)},p^{(v)},1,1,\eta_n\}$, $\Delta D_n=\{d^{(m)},d^{(v)},Y_n,Y_n,\delta_n\}$, $\Delta E_n=\{e^{(m)},e^{(v)},Z_n,Z_n,\epsilon_n\}$, $\Delta F_n=\{f^{(m)},f^{(v)},U^f(Y_n, Z_n),V^f(Y_n, Z_n),\phi_n\}$, and $\Delta G_n=\{g^{(m)},g^{(v)},U^g(Y_n, Z_n),V^g(Y_n, Z_n),\gamma_n\}$. Finally, the Milstein terms $\Delta M_n$ and $\Delta N_n$ in (\ref{BC}) read as:
\Eqn 
 \Delta  M_n    &=&   - \frac 14  d^{(v)} ( (\Delta  \delta_n)^2 - \Delta t) \nonumber \\ && - \frac \alpha 4\left[ f^{(v)} \frac {\partial V^f}{\partial Y_n}  ( (\Delta  \phi_n)^2 - \Delta t)\, -  g^{(v)} \frac {\partial V^g}{\partial Y_n}  ( (\Delta  \gamma_n)^2 - \Delta t)\,   \right]  , \nonumber \\
 \Delta  N_n    &=&  - \frac 14  e^{(v)} ( (\Delta  \epsilon_n)^2 - \Delta t) \nonumber \\ && +  \frac 1 4\left [ f^{(v)} \frac {\partial V^f}{\partial Z_n}  ( (\Delta  \phi_n)^2 - \Delta t)\, -  g^{(v)} \frac {\partial V^g}{\partial Z_n}  ( (\Delta  \gamma_n)^2 - \Delta t)\,  \right ]  . \qquad\label{BC3}
\Eeqn 
Also, we assume (as in the case without buffer) that $Y_n$ and $Z_n$ converge weakly towards the steady states $Y$ and $Z$, respectively, in the steady-state limit $n\to\infty$ defined previously.

The linear combination of the two equations (\ref{BC}) that eliminates the interconversion rates (except through the Milstein terms) yields the following conservation equation:
\Eq
 [Y_{n+1} +  \alpha \,  Z_{n+1}] = [Y_n +  \alpha \,  Z_n] +  [\Delta  P_n ] -  [\Delta D_n   + \alpha \Delta E_n]+ [\Delta M_n  + \alpha \Delta N_m ]  .  \label{BC4}
\Eeq
Taking the mean of both sides yields, in the steady state limit $n\to \infty$:
\Eq
p^{(m)} = d^{(m)} \, \E(Y)  + \alpha \, e^{(m)} \, \E(Z)   ,  \label{BC5}
\Eeq
which means that, in these limits, the production and the total degradation compensate each other on the average. Furthermore, taking the mean of  the square of both sides of Eq.(\ref{BC4}) gives, in the same limits, the following relation between variances and covariances:
\Eqn
  d^{(m)} \, \Var ( Y )+  \alpha^2 \, e^{(m)} \, \Var ( Z ) &=& \frac 12 \left ( p^{(v)} +  d^{(v)}    \E(Y) \right ) + \frac {\alpha^2} 2 \left ( e^{(v)}  \,  \E(Z) \right ) \nonumber \\ &&\hspace{0.1cm} -  \alpha \left ( d^{(m)} + e^{(m)} \right ) \Cov (Y,Z)  .  \label{BC6}
\Eeqn
Note that this equation does not depend explicitly on the type of interconversion between molecules of types $y$ and $z$. Note also that the instantaneous covariance  $\Cov(Y,Z)$ is equal to the delayed  covariance up to terms of the order of $\Delta t$, which vanish in the continuous time limit:
\Eq
\Cov(Y_n,Z_n)=\Cov(Y_{n+i},Z_n) +O(\Delta t) = \Cov(Y_n,Z_{n+i}) +O(\Delta t) ,
\Eeq
where $i$ is an integer.

To obtain  relations that involve the interconversion terms, consider again the two equations~(\ref{BC}). The computation of their mean and the mean of their squares yields the following additional relations in the same limits:
\Eqn 
  p^{(m)}  - d^{(m)} \, \E(Y) &=&   \alpha  \E(\Phi^{y \to z}) ~~= ~~  \alpha  e^{(m)} \, \E(Z)  , \label{BC7} \\
  d^{(m)}  \Var ( Y ) & = & \frac 12 \left ( p^{(v)} +  d^{(v)}  \E(Y) \right )  - \alpha \, \Cov (Y, \Phi^{y \to z})  + \frac {\alpha^2} 2 \Phi^{(v)}  , \nonumber \\
  e^{(m)}  \Var ( Z ) & = & \frac 12 \left ( e^{(v)}  \E(Z) \right )  -  \Cov (Z, \Phi^{z \to y})  + \frac 1 2 \Phi^{(v)}  , \nonumber \\
   (d^{(m)} + e^{(m)} )\Cov (Y,Z) &=& \Cov(Y,\Phi^{y \to z})+ \Cov( \alpha Z, \Phi^{z \to y}) - \alpha \Phi^{(v)}  , \nonumber
\Eeqn 
where the "mean" flux  $\Phi^{y \to z}$ from system $y$ to buffer $z$ in the steady-state limit is defined by: $\Phi^{y \to z}= -\, \Phi^{z \to y}=f^{(m)}U^f- g^{(m)}U^g $, and the flux "variance" by $\Phi^{(v)}=f^{(v)}  \E (V^f)  + g^{(v)}\E (V^g )  $. The first equation means that, in the steady-state limit, the average number of molecules entering system $y$ is equal to the number of molecules leaving it, and similarly for $z$. The other three equations relate $ \Var(Y)$, $ \Var(Z)$ and $\Cov(Y,Z)$  to the  "mean"  flux $\Phi^{y \to z}$ and the flux "variance" $\Phi^{(v)}$ between the system and the buffer.

Let us now compare the systems with and without a buffer. Consider therefore that $y$ and $\tilde y$ refer to the same type of molecules. Their degradation rates are thus identical. Moreover, to make things comparable, we need to impose an equal average number of molecules in the steady-state limit, {\it i.e.} $\E(Y)=\E(\tilde Y)$. This implies that the mean production rates are in general different to allow a sufficient production and compensate on the average for the molecules of type $y$ that enter the buffer. However, we assume that the production in both systems is due to the same external process and is thus of the same kind, so that the ratio of the variance and mean of the production rates are identical. More precisely, we set:
\Eq
 d^{(m)}  =  \tilde d^{(m)}   , \quad d^{(v)}  =  \tilde d^{(v)}  , \quad  \frac {p^{(v)}}{p^{(m)}}=\frac {\tilde p^{(v)}}{\tilde p^{(m)}} , \quad \E(Y)=\E(\tilde Y) = \frac {\tilde p^{(m)}} { d^{(m)}}   .  \label{C}
\Eeq
Inserting these relations into Eqs(\ref{BC7}) and using Eq.(\ref{B2_2}) yields:
\Eq
p^{(m)}-\tilde p^{(m)}=\alpha \, \E(\Phi^{y \to z})   ,  \label{C2}
\Eeq
and the following expression for the variance of the system without a buffer:
\Eq
\Var (\tilde Y)= \frac {\E(Y)} 2 \left ( \frac {p^{(v)}}{ p^{(m)}} + \frac {d^{(v)}}{d^{(m)}}\right  )       .  \label{C3}
\Eeq
The use of these relations and of Eqs(\ref{BC6}-\ref{BC7}) gives the difference between the variances for systems with and without a buffer :
\Eqn 
 &&d^{(m)} \left ( \Var ( \tilde Y ) - \Var ( Y )\right ) =   \alpha \, \Cov (Y, \Phi^{y \to z}) - \frac {\alpha^2} 2  \Phi^{(v)}  
   - \frac \alpha 2  \frac {p^{(v)}}{p^{(m)}} \E(\Phi^{y \to z})  \qquad  \label{C4} \\ 
 &&\qquad = \alpha ^2 e^{(m)} \left ( \Var ( Z ) - 
( \frac {e^{(v)}}{e^{(m)}} +  \frac {p^{(v)}}{\alpha p^{(m)}} ) 
 \frac{\E(Z)} 2 \right ) + \alpha \left ( d^{(m)} + e^{(m)}  \right ) \Cov (Y,Z)   \nonumber \\ 
&&\qquad  =  - \alpha^2 \, \Cov ( Z,  \Phi^{z \to y}) + \frac {\alpha ^2} 2 \, \Phi^{(v)} 
   - \frac \alpha 2  \frac {p^{(v)}}{p^{(m)}} \E(\Phi^{y \to z}) 
+ \alpha \left ( d^{(m)} + e^{(m)}  \right ) \Cov (Y,Z)  . \nonumber
\Eeqn 

First consider the case where the degradation rate of the buffer ($\Delta E_n$) vanishes. Eqs(\ref{C4}) then imply  that the  variance of the system without a buffer is larger than the variance of the system with a buffer if and only if $\Cov(Y,Z)$ is positive, {\it i.e.} if and only if $Y$ and $Z$ are positively correlated:
\Eq 
 \Delta E_n=0 : \quad \Var ( Y ) \le   \Var ( \tilde Y )  \iff   \Cov (Y,Z) \ge 0   . \label{C5}
\Eeq 
This condition may be considered as  satisfied in general: a buffer is indeed by definition positively correlated with the system, since its role is to absorb the possible overflow of molecules of the system, or, on the contrary, to provide it with molecules if it runs short. 

On the other hand, when $\Delta E_n$ does not vanish, the difference between the variances of the systems with and without a buffer is equal to  $\Cov(Y,Z)$ plus a term that depends on the mean and variance of $Z$ (Eqs(\ref{C4})):
\Eqn 
&&\qquad  \qquad \Var ( Y ) \le   \Var ( \tilde Y )  \iff     \label{C7}\\
&&  \left ( d^{(m)} + e^{(m)}  \right ) \Cov (Y,Z)   + \alpha e^{(m)} \left ( \Var ( Z ) 
- ( \frac {e^{(v)}}{e^{(m)}}  +  \frac {p^{(v)}}{\alpha p^{(m)}} ) \frac{\E(Z)} 2 \right )  \ge 0   . \nonumber
\Eeqn 
 This additional term may be positive or negative in general. It is positive in the case the Fano factor of $Z$ is of order one, and when the buffer's degradation gate 
 and the systems's production gate are not too noisy, {\it i.e.} when $e^{(v)}$ and  $p^{(v)}$ are not too much larger than $e^{(m)}$ and $p^{(m)}$. If these two reasonable conditions are satisfied and if $\Cov(Y,Z)$ is positive, we can again deduce that the variance of the system without a buffer is larger than that of the system with a buffer.
An equivalent condition is obtained from the second equation in Eqs.(\ref{C4}):
\Eq 
 \Var ( Y ) \le   \Var ( \tilde Y )  \iff    \Cov (Y, \Phi^{y \to z}) \ge \frac 12  \left (  \alpha \Phi^{(v)} 
  +   \frac {p^{(v)}}{p^{(m)}} \E(\Phi^{y \to z})  \right ). \label{C8}
\Eeq 
This condition is satisfied when the "mean" flux $\Phi^{y \to z}$ towards the buffer is positively correlated with $Y$ (which again is a reasonable assumption for a buffer) and when the flux "variance" $\Phi^{(v)} $ and "mean" $\E(\Phi^{y \to z}) $ are not too large; note that $\Phi^{(v)} $ is always positive but that $\E(\Phi^{y \to z}) $ may be positive or negative.

To illustrate that these conditions are indeed satisfied for quite general types of biological buffers, and thus that the presence of such buffers tends to reduce the variance of the number of molecules, we analyze in detail two different model systems.
 
\subsection{Model 1: active and nonactive states of a protein}
Consider the case where $y$ and $z$ correspond to the same protein in two different structural states: $y$ corresponds to the active state and $z$ to the inactive one. We then have:
\Eqn
\alpha=1  , &&U^f(Y_n,Z_n)=Y_n=V^f(Y_n,Z_n)   , \nonumber \\
 &&U^g(Y_n,Z_n)=Z_n =V^g(Y_n,Z_n)  . \label{Act1}
\Eeqn
From Eqs (\ref{BC7}) we can easily solve for $\E(Y)$, $\Var(Y)$, $\E(Z)$, $\Var(Z)$ and $\Cov(Y,Z)$ in terms of the system's  parameters: 
\Eqn
c \,\E(Y) &=& a \,  p^{(m)} 
 ,  \label{Act2} \\
c \, \E(Z) &=& f^{(m)} p^{(m)} 
 , \nonumber  \\
2 \, c \, (a+b) \, \Var(Y) 
& = &  (c+a^2  ) p^{(v)} + \frac {p^{(m)}}c \left (  a (c+a^2)    d^{(v)} 
+  g^{(m)2} f^{(m)} e^{(v)}\right.\nonumber\\
&&\hspace{1cm}\left.+(c+e^{(m)2} ) (a f^{(v)}+f^{(m)} g^{(v)}\right ) 
   , \nonumber \\
2\, c \, (a+b) \, \Var(Z) 
 &=&  f^{(m)2} p^{(v)} + \frac {p^{(m)}}c \left ( a f^{(m)2} d^{(v)}
+   ( c+b^2) f^{(m)} e^{(v)}\right.\nonumber\\
&&\hspace{1cm}\left.+ (c+  d^{(m)2}) 
 (a f^{(v)}+f^{(m)}g^{(v)} \right ) 
  , \nonumber \\
2\,c \, (a+b) \, \Cov(Y,Z) 
&=& a f^{(m)} p^{(v)}+\frac {p^{(m)}}c \left ( a^2 f^{(m)} d^{(v)} + b g^{(m)} f^{(m)} e^{(v)} \right.\nonumber\\
&&\hspace{1cm}\left.-  (a d^{(m)}+b e^{(m)})(a f^{(v)} + f^{(m)} g^{(v)}) \right ) , \nonumber 
\Eeqn
where
\Eqn
a&=& e^{(m)}+g^{(m)}
\, , \nonumber \\
b&=& d^{(m)}+f^{(m)}
\, , \nonumber \\
c&=&d^{(m)} e^{(m)} + e^{(m)} f^{(m)} + d^{(m)}g^{(m)} 
\, \nonumber.
\Eeqn
Using Eq(\ref{C3}), we  find that:
\Eqn
 &&\qquad \qquad\Var ( Y )  \le  \Var ( \tilde Y )  \iff   \label{Act3} \\
 &&   0 \le  c \, g^{(m)} \left(\frac{p^{(v)}} {p^{(m)}} -\frac{g^{(v)}} {g^{(m)}} \right)+e^{(m)} g^{(m)2} \left(\frac{d^{(v)}} {d^{(m)}} -\frac{e^{(v)}} {e^{(m)}} \right)
 \nonumber \\  
&&\quad +\,a(c+e^{(m)2}) \left(\frac{d^{(v)}} {d^{(m)}} -\frac{f^{(v)}} {f^{(m)}} \right)
+e^{(m)2} g^{(m)} \left(\frac{d^{(v)}} {d^{(m)}} -\frac{g^{(v)}} {g^{(m)}} \right) . \nonumber 
\Eeqn 
In the unperturbed birth-death case, where $d^{(v)} =d^{(m)} $, $e^{(v)} =e^{(m)} $, $p^{(v)} =p^{(m)} $, $f^{(v)} =f^{(m)} $ and $g^{(v)} =g^{(m)} $, we have the equality $\Var(Y)=\Var(\tilde Y)$.  More generally, this equality remains true when  all the parts of the process are equally noisy, in the sense that $\frac{p^{(v)}}{p^{(m)} }=\frac{d^{(v)}}{d^{(m)} }=\frac{e^{(v)}}{e^{(m)} }=\frac{f^{(v)}}{f^{(m)} }=\frac{g^{(v)}}{g^{(m)} }$.  In that case, we also note that $\Cov(Y,Z)=0$.

In contrast, the inequality  $\Var(Y)< \Var(\tilde Y)$ is satisfied if the buffer is less noisy than the main system, {\it i.e.} if $\frac{p^{(v)}}{p^{(m)} },\frac{d^{(v)}}{d^{(m)} }< \frac{e^{(v)}}{e^{(m)} },\frac{f^{(v)}}{f^{(m)} },\frac{g^{(v)}}{g^{(m)} }$. Conversely, if the buffer is more noisy than the main system, {\it i.e.} if $\frac{p^{(v)}}{p^{(m)} },\frac{d^{(v)}}{d^{(m)} }> \frac{e^{(v)}}{e^{(m)} },\frac{f^{(v)}}{f^{(m)} },\frac{g^{(v)}}{g^{(m)} }$,  we have the reversed inequality $\Var(Y)>\Var(\tilde Y)$. Note that the covariance $\Cov(Y,Z)$ is positive if the interconversion gates are not too noisy, and becomes negative otherwise.

\bigskip
Now consider a slightly different case where the proteins are produced in their inactive state, and become active in the "buffer" (the quotation marks indicate that it is no longer a true buffer). In this case we have to compare $\Var(Z)$ to $\Var(\tilde Y)$, with conditions that  differ from those given in Eq.(\ref{C}). These  are:
\Eqn
&& e^{(m)}  =  \tilde d^{(m)}   , \qquad e^{(v)}  =  \tilde d^{(v)}  , \qquad  \frac {p^{(v)}}{p^{(m)}}=\frac {\tilde p^{(v)}}{\tilde p^{(m)}} ,  \label{Act4}  \\
&& \E(Z)=\E(\tilde Y) = \frac {\tilde p^{(m)}} { e^{(m)}} 
  , \quad  
 \Var (\tilde Y)= \frac {\E(Z)} 2 \left ( \frac {p^{(v)}}{p^{(m)}} + \frac {e^{(v)}}{e^{(m)}}\right  )    . \nonumber 
\Eeqn
With these conditions and Eqs(\ref{Act2}), we obtain the following result:
\Eqn
&& \qquad \qquad \Var ( Z ) \le   \Var ( \tilde Y)  \iff   \label{Act5} \\
 && 0 \le  a (c+d^{(m)2}) (\frac{p^{(v)}} {p^{(m)}} -\frac{f^{(v)}} {f^{(m)}} )+d^{(m)} f^{(m)} g^{(m)} (\frac{e^{(v)}} {e^{(m)}} -\frac{d^{(v)}} {d^{(m)}} )
 \nonumber \\  
&&+\,d^{(m)}e^{(m)}f^{(m)} (\frac{p^{(v)}} {p^{(m)}} -\frac{d^{(v)}} {d^{(m)}} )
+ \quad g^{(m)}(c+d^{(m)2}) (\frac{e^{(v)}} {e^{(m)}} -\frac{g^{(v)}} {g^{(m)}} )
   .\ \nonumber  
\Eeqn 
Again, the equality $\Var(Z)=\Var(\tilde Y)$ is satisfied  in the unperturbed birth-death model and, more generally, when the  "buffer" and the main system are equally noisy.  In such a case we have $\Cov(Y,Z)=0$. The inequality  $\Var(Z)< \Var(\tilde Y)$ is satisfied when the "buffer" is noisier than the main system, while the reversed inequality $\Var(Z)> \Var(\tilde Y)$ holds when the main system is noisier than the "buffer".

\bigskip
Note that this system can also be viewed as modeling molecules that are located in two different cell compartments, or that are inside and outside the cell. 

\subsection{Model 2: protein monomers and dimers}

Consider the case where the molecules $y$ are protein monomers and  the molecules $z$ are homodimers formed of two molecules $y$. We then have:
\Eqn
\alpha=2 , & \qquad & U^f(Y_n,Z_n)=\frac 1 {2!}Y_n(Y_n-1)=V^f(Y_n,Z_n)  , \nonumber \\
&& U^g(Y_n,Z_n)=Z_n=V^g(Y_n,Z_n)  . \label{dim}
\Eeqn
Let us first assume that the buffer's degradation rate vanishes, {\it i.e.} $e^{(v)} =0=e^{(m)}$. Inserting relation (\ref{dim}) in Eqs(\ref{BC}), and taking the mean of these equations as well as the mean of their squares and of their product, we can obtain $\E(Y)$, $\E(Z)$, $\Var(Z)$ and $\Cov(Y,Z)$ as functions of the parameters, of $\Var(Y)$ and of the skewness $\kappa(Y)=\E \left ((Y-\E(Y))^3 \right )$. In this way, we get the relation:
\Eq
\left(A \E(Y)+B \right )\left (\Var (\tilde Y) - \Var( Y) \right ) + C \left (\kappa(\tilde Y)-\kappa(Y) \right )=D \E(Y)^2 + E \E(Y)
  , \label{dim1}
\Eeq
where
\Eqn
A&=& 4 f^{(m)} , \nonumber\\
B&=& 2  \left ( d^{(m)} + g^{(m)} -f^{(m)}(1 +\frac{f^{(v)}}{f^{(m)}}+\frac{g^{(v)}}{g^{(m)}})\right) ,\nonumber \\
C&=&2 f^{(m)} ,\nonumber \\
D&=& 2 f^{(m)}  \left (\frac{d^{(v)}}{d^{(m)}}+\frac{p^{(v)}}{p^{(m)}}-\frac{f^{(v)}}{f^{(m)}}-\frac{g^{(v)}}{g^{(m)}}\right ) , \nonumber  \\
E&=&f^{(m)}
\left(\left (2-\frac{d^{(v)}}{d^{(m)}}-\frac{p^{(v)}}{p^{(m)}}\right)
 \left( \frac{f^{(v)}}{f^{(m)}}+\frac{g^{(v)}}{g^{(m)}}\right)+
 \left (\frac{d^{(v)}}{d^{(m)}}-1\right) \left(\frac{d^{(v)}}{d^{(m)}}+\frac{p^{(v)}}{p^{(m)}}\right)\right) . \nonumber 
\Eeqn
The variance $\Var(\tilde Y)$ of the system without a buffer is given by Eq.(\ref{C3}). Its skewness is easy to compute by taking the mean of Eq.(\ref{B}) to the third power, which yields:
\Eq
 \kappa (\tilde Y)= \frac{d^{(v)}}{d^{(m)}} \Var(\tilde Y) . \label{dim3}
\Eeq
The skewness of the system without a buffer is thus proportional to $\Var ( \tilde Y)$.
Hence, for normal buffers, the skewness $\kappa(Y)$ of the system with a buffer can be assumed to be of the order of its variance $\Var ( Y)$. For $\E(Y)>>1$, we can thus focus on the term $\E(Y) (\Var(\tilde Y)-\Var (Y) )$ on the left-hand side of Eq.(\ref{dim1}) and on the term $\E(Y)^2$ on the right-hand side. This approximation gives:
\Eq
 \Var (\tilde Y) - \Var( Y)  \approx   \frac 12 \left (\frac{d^{(v)}}{d^{(m)}}+\frac{p^{(v)}}{p^{(m)}}-\frac{f^{(v)}}{f^{(m)}}-\frac{g^{(v)}}{g^{(m)}}\right ) \E(Y)
  . \label{dim4}
\Eeq
This equation means that, when $e^{(v)} =0=e^{(m)}$ and $\E(Y)>>1$, we have:
\Eqn
\Delta E_n=0 : \quad \Var(Y)<\Var(\tilde Y) &\iff&  \frac{d^{(v)}}{d^{(m)}}+\frac{p^{(v)}}{p^{(m)}}>\frac{f^{(v)}}{f^{(m)}}+\frac{g^{(v)}}{g^{(m)}}
\quad , \nonumber \\
\Var(Y)>\Var(\tilde Y) &\iff&  \frac{d^{(v)}}{d^{(m)}}+\frac{p^{(v)}}{p^{(m)}}<\frac{f^{(v)}}{f^{(m)}}+\frac{g^{(v)}}{g^{(m)}}
\quad . \label{dim5}
\Eeqn
The equality $ \Var(Y)=\Var(\tilde Y)$ is obtained in the unperturbed birth-death case or more generally when the parameters' mean and variance are equal ($ \frac{d^{(v)}}{d^{(m)}}=\frac{p^{(v)}}{p^{(m)}}=\frac{f^{(v)}}{f^{(m)}}=\frac{g^{(v)}}{g^{(m)}}$).

\bigskip
To check the validity of the assumptions made to obtain these relations, we performed some numerical simulations using the R package, which are summarized in Table \ref{Tab:ici1}. We chose as time step $\Delta t=0.1$, as initial conditions $Y_0=10=Z_0$, as number of time steps $N=10,000$, and made $10,000$ runs for the estimation of the mean and variance. As expected, in the unperturbed birth-death case, the Fano factor $\Var(Y)/\E(Y)$ remains roughly the same in the presence or absence of the buffer. It is reduced when the buffer is less noisy and increased when it is more noisy than the main system. An example, with parameters describing an unperturbed birth-death process,  is depicted in Figs \ref{fig2a} and  \ref{fig3a}. The Fano factor is in this case very similar (1.07 and 1.00) with or without  a buffer. Note that  the steady state is reached much faster  in the absence of a buffer.

\bigskip

Consider now the general case when the buffer's degradation rate does not vanish, thus $e^{(v)}\ne0\ne e^{(m)}$. Using the same procedure as above yields:
\Eqn
&&\left(A'\E(Y)+B'\right) \left (\Var (\tilde Y) -\Var( Y) \right ) + C' \left (\kappa(\tilde Y)-\kappa(Y) \right )\nonumber\\ 
&&\qquad \qquad+F' \left ( \E(Z) -\Var ( Z) \right )= D'\E(Y)^2 + E' \E(Y) , \qquad
\Eeqn
where
\Eqn
A'&&= 2 d^{(m)} g^{(m)} A + 8 e^{(m)} f^{(m)} \left( d^{(m)} +e^{(m)} +g^{(m)}\right) 
, \nonumber \\
B'&&= 2 d^{(m)} g^{(m)} B 
+4 e^{(m)} d^{(m)} \left( d^{(m)} +e^{(m)}  +2 g^{(m)}\right) \nonumber \\
&&-2 e^{(m)} f^{(m)}\left(d^{(m)} +e^{(m)} \right)\left( 2+2 f+p \right)  + e^{(m)}g^{(m)}\left(2 f+p+2g+2e-2\right)
, \nonumber \\
C' &&= 2 d^{(m)} g^{(m)} C+4  e^{(m)} f^{(m)} \left( d^{(m)} +e^{(m)} + g^{(m)} \right) ,\nonumber \\
D'&&= 2 d^{(m)} g^{(m)} D \nonumber \\
&&+ 2 e^{(m)} f^{(m)} \left( \left( d^{(m)} +e^{(m)} + g^{(m)} \right )\left (p+2d-2f \right)+ 2 g^{(m)}\left(2 -e-g \right)\right)
 ,\nonumber \\
E'&&=2 d^{(m)} g^{(m)} E+2 e^{(m)} f^{(m)} g^{(m)}\left ( 2- e-g\right)\left( d+p -2\right)\nonumber \\
&&+e^{(m)}f^{(m)}\left( d^{(m)} +e^{(m)} + g^{(m)} \right )
\left(2d^2-p^2+\left(2-p\right)\left(2f-d \right)- 2 f d\right)  , \nonumber \\
F'&&=16 e^{(m)}g^{(m)}\left(e^{(m)}+g^{(m)}\right ) ,
\nonumber
\Eeqn
with $p=\frac{p^{(v)}}{p^{(m)}}$, $d=\frac{d^{(v)}}{d^{(m)}}$, $e=\frac{e^{(v)}}{e^{(m)}}$, $f=\frac{f^{(v)}}{f^{(m)}}$, $g=\frac{g^{(v)}}{g^{(m)}}$. Again, as argued before, we can disregard the term involving the skewness $\kappa(Y)$ when $\E(Y)>>1$ since it should in general be of the order of $\Var(Y)$. Furthermore, we also expect $\Var(Z)\sim\E(Z)$ for normal buffers. If we moreover focus on parameters that yield a similar amount of molecules in the system and in the buffer, so that $\E(Z)$ and $\E(Y)$ are of the same order, the term $\E(Z)-\Var(Z)$ can be disregarded too. The dominating terms for $\E(Y)>>1$ are thus:
\Eq
 A''\left (\Var (\tilde Y) - \Var( Y) \right ) \approx D''\E(Y), \label{dim8}
\Eeq
where
\Eqn
A''&=& 4 \left( d^{(m)} +e^{(m)} \right) \left( e^{(m)} +g^{(m)} \right) 
\quad, \nonumber \\
D''&=& e^{(m)}  \left( d^{(m)} +e^{(m)} +g^{(m)} \right) \frac{p^{(v)}}{p^{(m)}} 
+ 2 \left( d^{(m)} +e^{(m)} \right) \left( e^{(m)} +g^{(m)} \right) \left(\frac{d^{(v)}}{d^{(m)}} -\frac{f^{(v)}}{f^{(m)}} \right)
\nonumber \\
&&+ 2 g^{(m)}e^{(m)} \left(2-\frac{e^{(v)}}{e^{(m)}} -\frac{g^{(v)}}{g^{(m)}} \right)
+ 2 g^{(m)}d^{(m)} \left(\frac{p^{(v)}}{p^{(m)}} -\frac{g^{(v)}}{g^{(m)}} \right)
 \quad.
 \label{dim9}
\Eeqn
Thus, as $A''$ is always positive, we deduce that the noise is reduced in the system by the presence of the buffer if and only if $D''$ is positive:
 \Eqn
\Var(Y)<\Var(\tilde Y) &\iff&  D''>0 , \nonumber \\
\Var(Y)>\Var(\tilde Y) &\iff&  D''<0  . \label{dim10}
\Eeqn
It is easy to see from Eq.(\ref{dim9}) that the noise in the main system is always reduced by the presence of a buffer in the unperturbed birth-death case or in the more general case where the parameters' means and variances  are equal, except when the buffer's degradation rate vanishes. In the latter case, the noise of the main system remains the same whether or not a buffer is present, in agreement with Eq.(\ref{dim5}). 

More generally, the noise of the main system is reduced by the presence of the buffer when the buffer is less noisy than the main system (${d^{(v)}}/{d^{(m)}}$, ${p^{(v)}}/{p^{(m)}}>{f^{(v)}}/{f^{(m)}}$, ${g^{(v)}}/{g^{(m)}}$, ${e^{(v)}}/{e^{(m)}}$). This remains true even if the buffer is somewhat more noisy than the main system  provided that $e^{(m)}\ne0$. It is only when the buffer becomes too noisy that the noise rate in the main system starts to increase  (${d^{(v)}}/{d^{(m)}},{p^{(v)}}/{p^{(m)}}<<{f^{(v)}}/{f^{(m)}},{g^{(v)}}/{g^{(m)}},{e^{(v)}}/{e^{(m)}}$).

\bigskip
This analysis is again based on a number of hypotheses. To check their validity, we made a series of numerical simulations, summarized in Table \ref{Tab:ici}. We chose again $\Delta t=0.1$, $Y_0=10=Z_0$, $N=10,000$ time steps, and $10,000$ runs for mean and variance estimation. As expected from the analytical development, we find that  the variance of the main system is increased by the presence of the buffer only when the variance parameters involving the buffer ({\it i.e.} $e^{(v)} $, $f^{(v)}$ and $g^{(v)}$) are much larger than their mean  (hence when the buffer is very "bad"). In all other cases, the variance is decreased by the presence of the buffer. In the unperturbed birth-death case, the noise reduction that is reached with the tested parameters  amounts to more than 20\%. Even larger noise reductions are obtained when the buffer is less noisy than the main system.

Three examples of trajectories with the associated probability densities at the steady state are depicted in Figs \ref{fig2} and \ref{fig3}. The first (Figs \ref{fig2b} and  \ref{fig3b}) corresponds to an unperturbed birth-death process, in which the Fano factor is reduced from  1 to 0.82 upon addition of a buffer. In Figs \ref{fig2c} and \ref{fig3c}, the interconversion gates are less noisy than usual ($\frac{f^{(v)}}{f^{(m)}}=\frac 12=\frac{g^{(v)}}{g^{(m)}}$) and the Fano factor is even more reduced by the buffer: from 1.00 to 0.56. In contrast, with very noisy interconversion rates ($\frac{f^{(v)}}{f^{(m)}}=2=\frac{g^{(v)}}{g^{(m)}}$), the Fano factor is increased from 1.00 to  1.39, as depicted in Figs \ref{fig2d} and \ref{fig3d} .

Two further observations can be made from Fig. \ref{fig2}. First, the number of molecules varies much more from one time step to the next in the presence of a buffer, even when approaching the steady state -- at least for the parameter values tested.  When no buffer is present,  the time variations in each particular trajectory seem more limited, even though the variance over the different  trajectories is sometimes much larger. Second,  the steady state is approached much faster in the absence of a buffer when $e^{(m)}=0=e^{(v)}$, whereas it is approached much faster in the presence of a buffer whose molecules can degrade ($e^{(m)}\ne0\ne e^{(v)}$). Further analyses are necessary to figure out whether these two  observations are general or instead  are specific to the parameter values tested.

\section{Two connected systems}
Now consider the slightly more complex system illustrated in Fig.\ref{fig1}(b), where the "buffer" has a non-vanishing production rate $\Delta Q_n$.  The two subsystems are thus perfectly symmetric, and we investigate whether their connection leads to a reduction of their respective variances. This system is modeled by the following coupled discretized SDEs:
 \Eqn 
 Y_{n+1} &=& Y_n +  \Delta  P_n -  \Delta D_n  - \alpha \, [\Delta F_n - \Delta G_n ] +\Delta M_n  ,\nonumber \\
 Z_{n+1} &=& Z_n    +  \Delta  Q_n -  \Delta E_n + [ \Delta F_n - \Delta G_n ] + \Delta N_n  . \label{K}
\Eeqn
Using the same approach as in the previous section, we compare the variances $\Var(Y)$ and $\Var(Z)$ of the connected systems to the variances $\Var(\tilde Y)$ and $\Var(\tilde Z)$ of the unconnected systems. We set thus, similarly to Eq.(\ref{C}):
\Eqn
 d^{(m)}  =  \tilde d^{(m)}   ,& \quad d^{(v)}  =  \tilde d^{(v)}  , \quad  \frac {p^{(v)}}{p^{(m)}}=\frac {\tilde p^{(v)}}{\tilde p^{(m)}} ,& \quad \E(Y)=\E(\tilde Y) = \frac {\tilde p^{(m)}} { d^{(m)}}   ,  \nonumber \\
 e^{(m)}  =  \tilde e^{(m)}   ,& \quad e^{(v)}  =  \tilde e^{(v)}  , \quad  \frac {q^{(v)}}{q^{(m)}}=\frac {\tilde q^{(v)}}{\tilde q^{(m)}} ,& \quad \E(Z)=\E(\tilde Z) = \frac {\tilde q^{(m)}} { e^{(m)}}   .  \label{K1}
\Eeqn
The variances of the unconnected systems $\tilde y$ and  $\tilde z$ read as:
\Eq
\Var (\tilde Y)= \frac {\E(Y)} 2 \left ( \frac {p^{(v)}}{ p^{(m)}} + \frac {d^{(v)}}{d^{(m)}}\right  )  \quad {\rm and }\quad \Var (\tilde Z)= \frac {\E(Z)} 2 \left ( \frac {q^{(v)}}{ q^{(m)}} + \frac {e^{(v)}}{e^{(m)}}\right  )   .  \label{K2}
\Eeq
The comparison between the variances of the connected and unconnected systems leads to the following relations, which are generalizations of Eqs(\ref{C4}):
\Eqn 
 & &d^{(m)}  \Var ( Y )+ \alpha ^2 e^{(m)} \Var ( Z ) \le   d^{(m)} \Var ( \tilde Y ) +  \alpha ^2 e^{(m)}  \Var ( \tilde Z ) \nonumber \\ 
 &&\qquad \qquad  \iff    \Cov (Y,Z) \ge 
 \frac1{2(d^{(m)}+e^{(m)})} \left (\frac {p^{(v)}}{p^{(m)}} - \frac {\alpha q^{(v)}}{ q^{(m)}}  \right )  \E(\Phi^{y \to z})  ,  \nonumber \\ 
  &&\Var ( Y ) \le   \Var ( \tilde Y )  \iff    \Cov (Y, \Phi^{y \to z}) \ge
  \frac 12  \left (  \alpha \Phi^{(v)} 
   +   \frac {p^{(v)}}{p^{(m)}} \E(\Phi^{y \to z}) \right ) ,  \nonumber \\ 
 &&\Var ( Z ) \le  \Var ( \tilde Z )  \iff    \Cov ( Z,  \Phi^{z \to y}) \ge 
 \frac 12  \left ( \Phi^{(v)} 
 +   \frac {q^{(v)}}{ q^{(m)}} \E(\Phi^{z \to y})  \right ) .\qquad\quad \label{K3}
\Eeqn 
We thus find that the (weighted) sum of the variances of the two systems is reduced upon connection if and only if  the correlation between their respective numbers of molecules is larger than a quantity that is proportional to the "mean" flux from the main system to the buffer; note that both the "mean" flux and the proportionality factor may be positive or negative. The variance of each of the subsystems is reduced if and only if the correlation between its number of molecules and the "mean" flux  towards the other system is larger than the  flux "variance" and "mean". 

The behavior of two positively correlated interacting systems is illustrated in Figs \ref{fig4} and  \ref{fig5}. The interconversion gates are considered to be the same as in the dimer case (Eq. (\ref{dim})); the  difference is that the "buffer" is now a true system in which molecules can be produced directly, independently of the other system ($q^{(m)}\ne0\ne q^{(v)}$). The parameters tested describe an unperturbed birth-death process. Both systems considered separately have a Fano factor of one, and this factor decreases upon connection (0.85 and 0.91). We have thus the expected decrease in noise upon connection of two systems whose molecules are positively correlated.

\section{Conclusion}

In this paper, we demonstrated analytically that the variance of the number of molecules in a system is decreased if it is connected to a particular kind of buffer. The comparison is performed upon imposing an equal mean number of molecules in the main system in the presence and absence of the buffer. The conditions that such buffers must {generally} satisfy are the following:
\begin{itemize}
\item The system and the buffer must be positively correlated. This means that the number of molecules in the system and in the buffer must be simultaneously higher or smaller than their respective means. It amounts to requiring that the buffer system  acts as a true buffer, which absorbs the excess of molecules produced in the system, or corrects its deficit. 
\item The Fano factor of the buffer must be of order one and its degradation gate must not be too noisy. This puts some reasonable constraints on the quality of the buffer.  Also, the system's production rate must not be too noisy.
\end{itemize}
Note that the first condition is sufficient when the buffer's degradation rate vanishes. These two conditions are equivalent to the following:
\begin{itemize}
\item The "mean" flux of molecules from the main system towards the buffer must be positively correlated with the number of molecules in the main system. Again this is a reasonable assumption for any bufferlike system. 
\item  The flux "variance" between the main system and the buffer must not be too large. This also is a reasonable assumption: if we connect a system with a buffer through a  highly noisy gate, we cannot expect it to reduce the noise in the main system. Furthermore, the "mean" flux towards the buffer must not be too positive.
\end{itemize}
These conditions are actually intuitive but cannot be proven analytically using the master equation formalism. The discrete SDE-based approach used in this paper has allowed us to  achieve this goal. 

We tested these general results in two explicit cases. In the first, the main system contains proteins in their active state and the buffer proteins in their inactive state, or conversely. The system-buffer conversion terms are in this case linear in the number of molecules, and the system of SDEs can be solved exactly in terms of the parameter values. For unperturbed birth-death processes, the system-buffer correlation vanishes and the noise is the same whether a buffer is present or not. When external perturbations modify the simple birth-death process, we found that the noise in the main system is decreased upon addition of a buffer that is less noisy than the main system, and increased otherwise. Note that this model can also be viewed as representing a molecule which, for example, moves from one cell compartment to another or goes from the cytoplasm to the nucleus. 

In the second test case,  the main system contains protein monomers and the buffer homodimers. The conversion terms between the main system and the buffer are in this case non-linear, and the system cannot be solved analytically unless some hypotheses are made about the skewness of the probability distributions and about the number of molecules in the main system compared to the buffer. With these assumptions, we found that the noise in the main system is reduced upon interaction with the buffer even in the unperturbed birth-death model. Higher levels of noise reduction are reached if the system is perturbed in such a way that the buffer is less noisy than the main system. The noise is seen to increase only when the buffer is much noisier than the main system. The validity of the hypotheses is supported by a series of numerical simulations. The noise reduction reaches 20\% for the tested parameter values in the case of an unperturbed birth-death process, and almost 50\% for a perturbed birth-death process.

Furthermore, we investigated analytically the more general case where two systems are connected, each containing molecules that are produced and degraded. The "buffer" and the main system are here considered on the same footing. We found the general result that the (weighted) sum of the variances of the number of molecules in the two systems is reduced upon connection, if their covariance is (sufficiently) positive; more precisely, it must be larger than a (positive or negative) quantity involving the "mean" flux between the systems. Focusing on one of the systems, the noise is seen to be reduced if and only if the covariance between the number of molecules and their "mean" flux  towards the other system is larger than a quantity involving the flux "variance" and "mean". Here again, the results are intuitive: systems have their noise level decreased upon interconnection when the molecules they contain are positively correlated.

The significance of the degrees of freedom encoded in the mean and variance of the parameters, generically denoted as $b^{(m)}$ and $b^{(v)}$, becomes clear at this point.  When an individual system unconnected to any other system contains molecules that are produced and degraded, it may be described by a simple unperturbed birth-death process. We  have in this case the equality $b^{(m)}=b^{(v)}$ and the number of molecules in the system follows a Poisson-type distribution with a Fano factor equal to one. When the system is connected to other systems, the distribution is in general no longer of Poissonian type and the Fano factor is either larger or lower than one. This non-Poisson behavior of a system due to its interaction with other systems can be recovered by considering the system as being unconnected but using effective production and degradation parameters  satisfying $b^{(m)} \ne b^{(v)}$, which encode the effect of the other systems without considering them explicitly. 

The perturbed birth-death processes characterized by $b^{(m)}\ne b^{(v)}$ that we analyzed in this paper can thus be viewed as representing systems whose noise is either increased by the interactions with other systems ($b^{(m)}<b^{(v)}$) or decreased ($b^{(m)}>b^{(v)}$).
For instance, the conversion of a protein from its inactive to its active state or its migration from one cell compartment to another usually occurs through binding with a ligand, which was not explicitly taken into account in the model. Since the number of ligands and the number of proteins are generally positively correlated, we may expect this protein-ligand interaction to cause an effective reduction of the variances of the parameters, and thus in particular $b^{(m)}>b^{(v)}$ for the system-buffer interconversion parameters. This would imply the noise reduction of the protein's active state, characterized by a Fano factor $< 1$. This issue will be further developed in future work.

This work opens many other interesting perspectives. The first consists in considering more complex cases -- for example cascades of interacting systems, which tend to better approximate real biological systems -- and study their effect on the noise level. Another perspective is to study the impact of a buffer on the time needed by the system to reach its steady state. Indeed, the simulations that we performed with a buffer having a non-vanishing degradation rate (Figs \ref{fig2b}-\ref{fig2d}) suggest that this time is much shorter in the presence of a buffer. If true, this complexification would  constitute an additional advantage for biological systems.

We thus conclude that connecting two systems or a system and a buffer tends to limit the overall noise in the case they are positively correlated. This result has clear implications in  cellular and molecular biology, since these contain a lot of systems that act cooperatively. Moreover, we saw that in the case of non-linear interactions, such as dimer formation from two monomers, the noise reduction is more pronounced. Strikingly again, many biological subsystems interact non-linearly. These findings suggest that a reason why biological systems interact and tend towards higher complexity across evolution is to reduce noise  and hence gain in predictable and robust behavior.

\bigskip
 
\section*{Acknowledgments}
We thank the Belgian fund for Scientific Research (FNRS) for support; JA is Postdoctoral Researcher and MR is Research Director at the FNRS.

\newpage
\section{References}

\bibliographystyle{plainnat}

\newpage
\section*{Tables}

\begin{table}[!h]
\centering
{\begin{tabular}{|c|c|c|}\hline 
& $\frac{\Var( \tilde Y)}{ \E(\tilde Y)}-\frac{\Var(Y)}{ \E(Y)}$ & number  of tests   \\ [1.5ex]
\hline
$\frac{d^{(v)}}{d^{(m)}}+\frac{p^{(v)}}{p^{(m)}}=\frac{f^{(v)}}{f^{(m)}}+\frac{g^{(v)}}{g^{(m)}}$& [-0.02, 0.03] &  16\\[0.5ex]
$\frac{d^{(v)}}{d^{(m)}}+\frac{p^{(v)}}{p^{(m)}}>\frac{f^{(v)}}{f^{(m)}}+\frac{g^{(v)}}{g^{(m)}}$&  [0.10, 0.46] &48\\[0.5ex]
$\frac{d^{(v)}}{d^{(m)}}+\frac{p^{(v)}}{p^{(m)}}<\frac{f^{(v)}}{f^{(m)}}+\frac{g^{(v)}}{g^{(m)}}$& [-0.90, -0.12] &80\\[0.5ex]
\hline
\end{tabular}
\caption{Numeric evaluation of change of the Fano factor upon protein dimerization in the case $e^{(v)}=0=e^{(m)}$. The parameters are equal to: $p^{(m)}=\{100,200\}$, $d^{(m)}=\{0.005,0.01\}$,  $f^{(m)}=\{5\,10^{-6},1\, 10^{-5}\}$,  $g^{(m)}=\{0.05,0.1\}$, with the relations: $\frac{p^{(v)}}{p^{(m)}}=1=\frac{d^{(v)}}{d^{(m)}}$, and $\frac{f^{(v)}}{f^{(m)}}=\{\frac 12,1,2\}=\frac{g^{(v)}}{g^{(m)}}$.  \label{Tab:ici1}}}
\end{table}

\bigskip

\begin{table}[!h]
\centering
{\begin{tabular}{|c|c|c|c|c|c|c|}\hline 
$\frac{p^{(v)}}{p^{(m)}}$&$\frac{d^{(v)}}{d^{(m)}}$&$\frac{e^{(v)}}{e^{(m)}}$&$\frac{f^{(v)}}{f^{(m)}}$&$\frac{g^{(v)}}{g^{(m)}}$& $\frac{\Var( \tilde Y)}{ \E(\tilde Y)}-\frac{\Var(Y)}{ \E(Y)}$ & number  of tests   \\ [1.5ex]
\hline
1&1&1&1&1 & [0.13, 0.22] &  32\\[0.5ex]
1&1&$\{\frac 12,1,2\}$& $\{\frac 12,1\}$&$\{\frac 12,1,2\}$& [0.05, 0.47] &  544\\[0.5ex]
1&1&$\{\frac 12,1,2\}$&$2$&$\{\frac 12,1,2\}$& [-0.46, -0.19] & 288\\[0.5ex]
\hline
\end{tabular}
\caption{Numeric evaluation of the change of the Fano factor upon protein dimerization when  $e^{(v)}\ne 0\ne e^{(m)}$. The parameters are equal to: $p^{(m)}=\{100,200\}$, $d^{(m)}=\{0.05,0.1\}=e^{(m)}$, $f^{(m)}=\{0.0005,0.001\}$, and $g^{(m)}=\{0.005,0.01\}$. The parameter values used in the first of row are excluded from the other rows. \label{Tab:ici}}}
\end{table}

\newpage
\section*{Figures}

\begin{figure}[!ht]
\begin{center}
\begin{tikzpicture}

\draw[very thick] (1,1) rectangle (3,3);
\draw (2,2) node {$\tilde Y$};
\draw (2,-0.5) node {$(a)$};
\draw[very thick]  (1.2,2) -- (0.2,2) ;
\draw[-triangle 60 reversed, thick]  (1.0,2)  -- (0.5,2) node[below=1mm] {$\tilde P$};
\draw[very thick]  (2,1.2) -- (2,0.2) ;
\draw[-triangle 60, thick]  (2,1) -- (2,0.5)node[right] {$\tilde D$};

\draw[very thick] (7,1) rectangle (9,3);
\draw (8,2) node {$ Y$};
\draw (9.5,-0.5) node {$(b)$};
\draw[very thick]  (7.2,2) -- (6.2,2) ;
\draw[-triangle 60 reversed, thick]  (7.0,2)  -- (6.5,2) node[below=1mm] {$P$};
\draw[very thick]  (8,1.2) -- (8,0.2) ;
\draw[-triangle 60, thick]  (8,1) -- (8,0.5)node[right] {$ D$};

\draw[very thick] (10,1) rectangle (12,3);
\draw (11,2) node {$Z$};
\draw[very thick, dashed]  (11.8,2)  -- (12.8,2) ;
\draw[-triangle 60 reversed, dashed, thick]  (12.4,2)  -- (12.5,2)  node[below=1mm] {$Q$};
\draw[very thick,dashed]  (11,1.2) -- (11,0.2) ;
\draw[-triangle 60, thick,dashed]  (11,1) -- (11,0.5)node[right] {$E$};

\draw[-left to, very thick]  (8.7,2.1)  -- (10.3,2.1);
\draw[-left to, very thick]  (10.3,1.9) -- (8.7,1.9) ;
\draw (9.5, 1.5) node {$F$};

\end{tikzpicture}
\caption{\textbf{Representation of a system without a buffer $(a)$, and  with a buffer $(b)$.}} \label{fig1}
\end{center}
\end{figure}

\begin{figure}[!H]
\subfigure[]{\label{fig2a}}{\centerline{\includegraphics[scale=0.7]{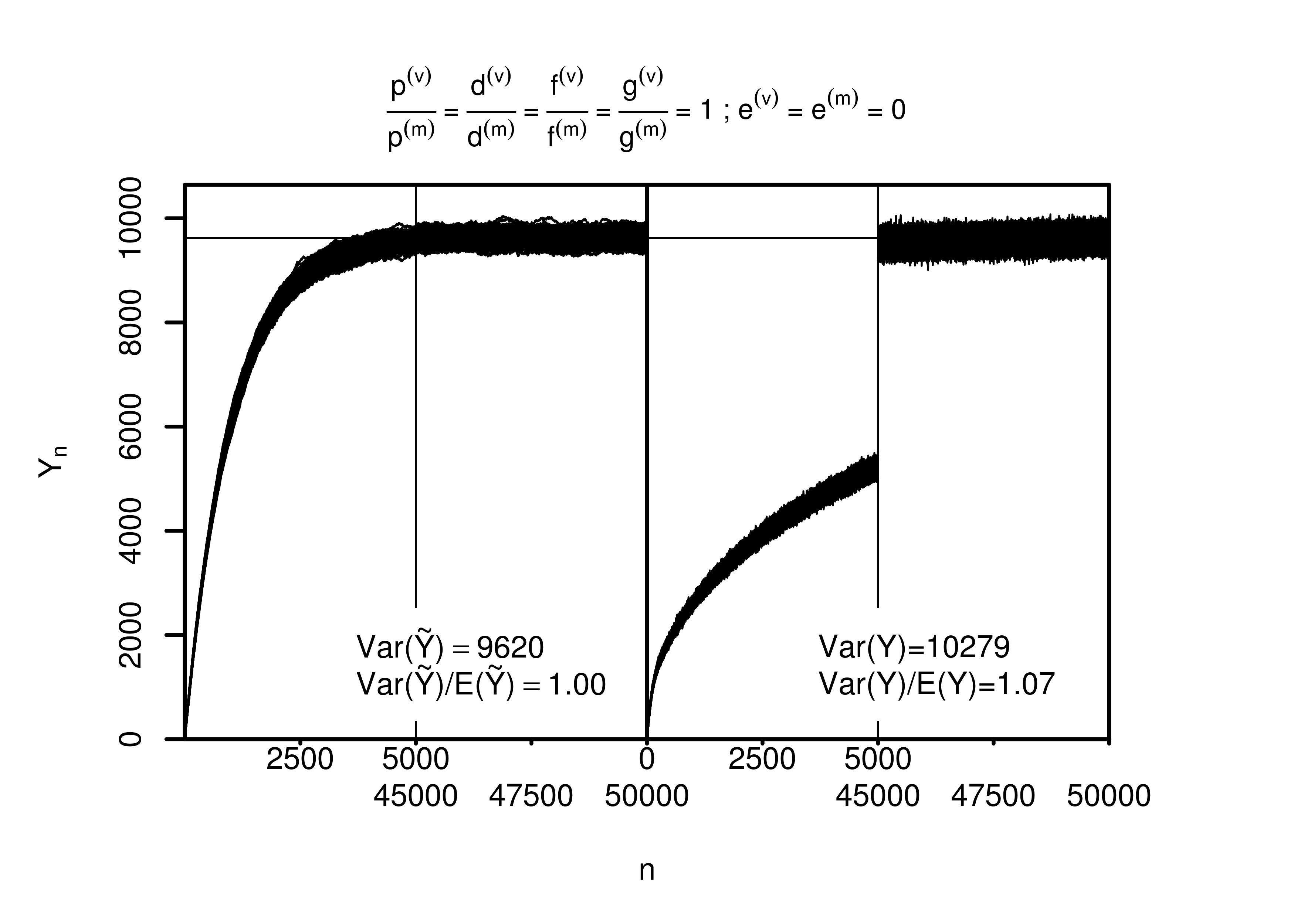}}}
\subfigure[]{\label{fig2b}}{\centerline{\includegraphics[scale=0.7]{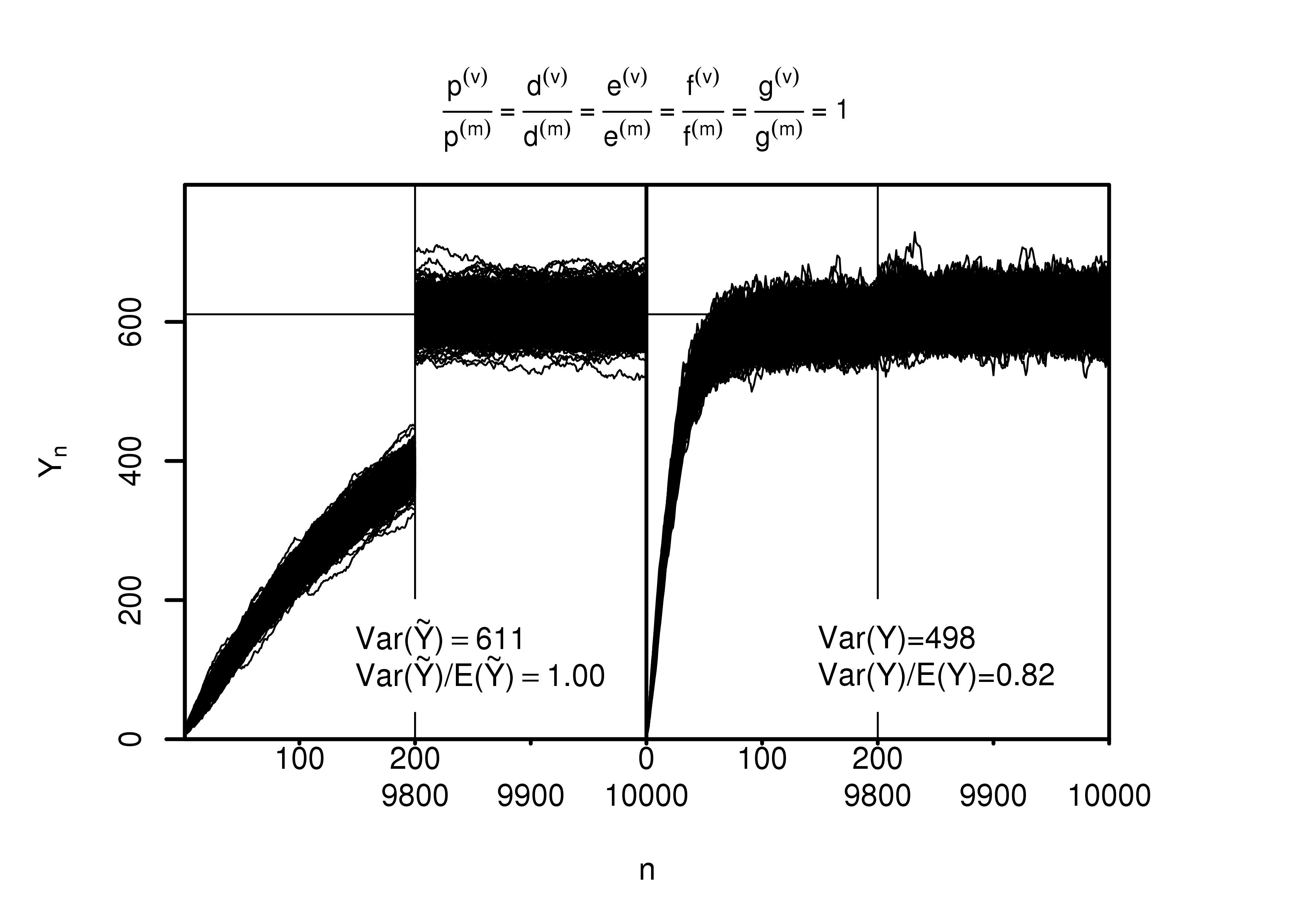}}}
\end{figure}
\begin{figure}[!ht]
\subfigure[]{\label{fig2c}}{\centerline{\includegraphics[scale=0.7]{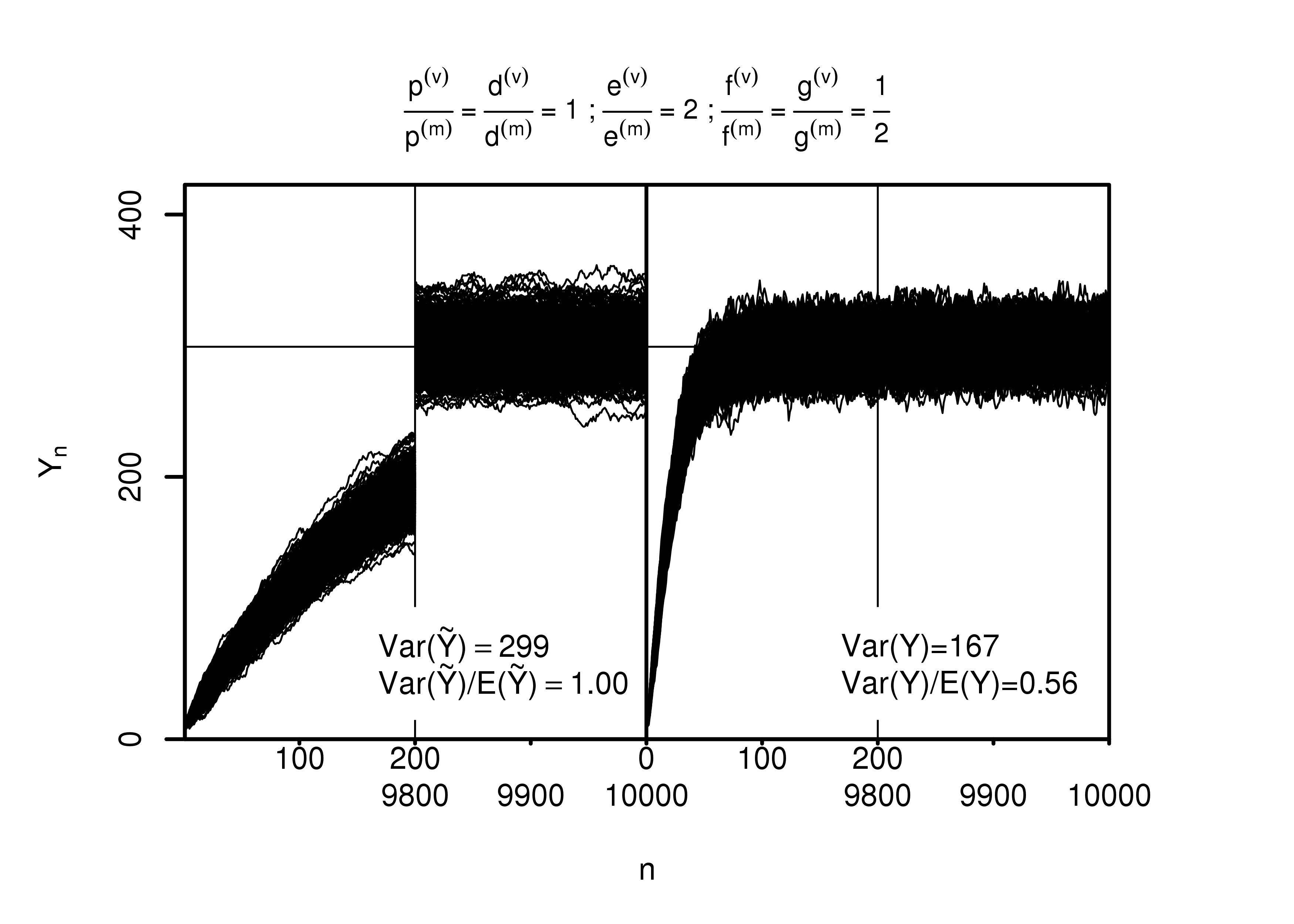}}}
\subfigure[]{\label{fig2d}}{\centerline{\includegraphics[scale=0.7]{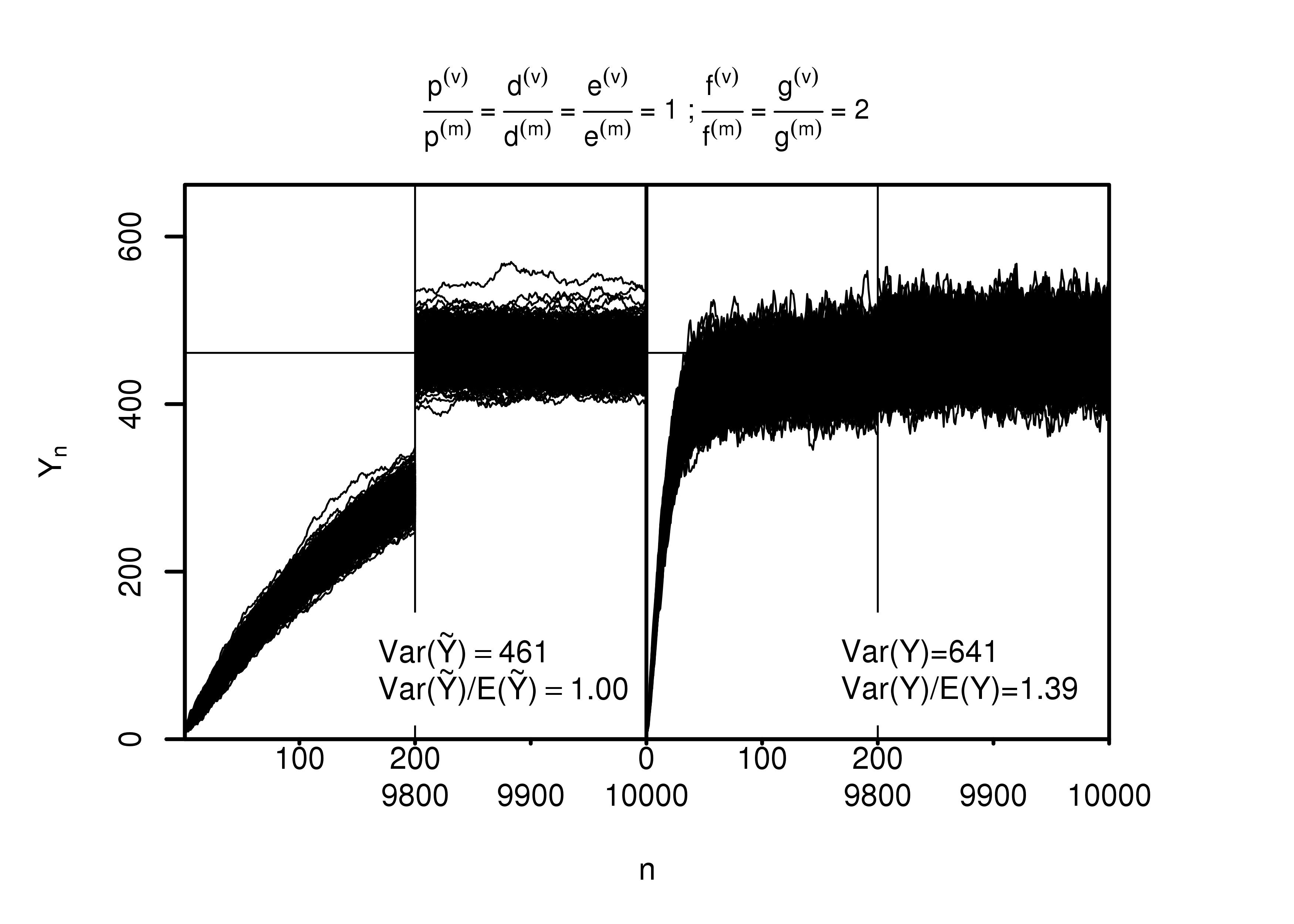}}}
\end{figure}
\begin{figure}[!ht]
\caption{\textbf{Numeric simulation of the number of molecules in a system without (left) and with (right) a buffer, as a function of time ($n$), in the case of protein dimerization.} The horizontal line represents the mean number of molecules in the steady state, which is imposed to be equal in the system with and without a buffer  ($\E(\tilde Y)=\E(Y)$).  (a) Unperturbed birth-death process with no degradation in the buffer: $e^{(v)}=0=e^{(m)}$, $p^{(m)}=100=p^{(v)}$, $d^{(m)}=0.01=d^{(v)}$, $f^{(m)}=5.10^{-5}=f^{(v)}$, $g^{(m)}=5.10^{-2}=g^{(v)}$; (b) Unperturbed birth-death process with degradation in the buffer:  $p^{(m)}=200=p^{(v)}$, $d^{(m)}=0.05=d^{(v)}$, $f^{(m)}=5.10^{-4}=f^{(v)}$, $g^{(m)}=5.10^{-3}=g^{(v)}$, and $e^{(m)}=0.05=e^{(v)}$; (c) Perturbed birth-death process with  little noisy interconversion gates:  $p^{(m)}=100=p^{(v)}$, $d^{(m)}=0.05=d^{(v)}$, $f^{(m)}=0.001=2 f^{(v)}$, $g^{(m)}=0.005=2 g^{(v)}$, and $e^{(m)}=0.1=\frac 12 e^{(v)}$; (d) Perturbed birth-death process with a noisy buffer: $p^{(m)}=200=p^{(v)}$, $d^{(m)}=0.05=d^{(v)}$, $f^{(m)}=0.001=\frac 12 f^{(v)}$, $g^{(m)}=0.1=\frac 12 g^{(v)}$, and $e^{(m)}=0.05=e^{(v)}$. The production parameter of the system without a buffer (${\tilde p^{(m)}}$) has been determined so as to have an equal mean number of molecules in the main system in the presence or absence of the buffer. The time step is $\Delta t = 0.1$, the initial conditions $Y_0 = 10 = Z_0$, the number of time steps $N = 50,000$ in (a) and $N = 10,000$ in (b-d), and the number of runs $R=10,000$. Only the first and last iterations are shown ($n\le 5,000$ and $n\ge 45,000$  in (a) and $n\le 200$ and $n\ge 9,800$), and  500 trajectories are drawn.}\label{fig2}
\end{figure}

\begin{figure}[!H]
\subfigure[]{\label{fig3a}}{\centerline{\includegraphics[scale=0.7]{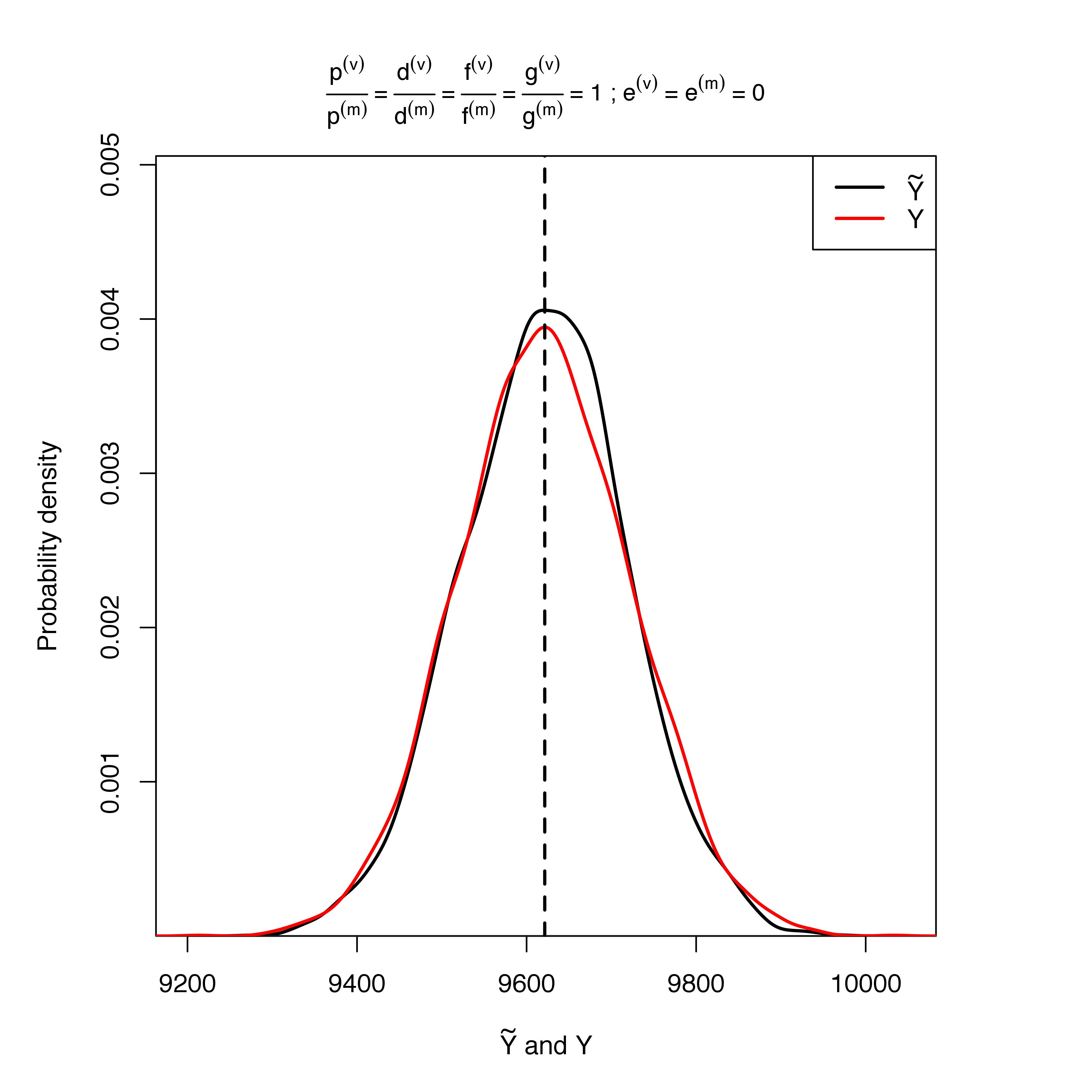}}}
\end{figure}
\begin{figure}[!ht]
\subfigure[]{\label{fig3b}}{\centerline{\includegraphics[scale=0.7]{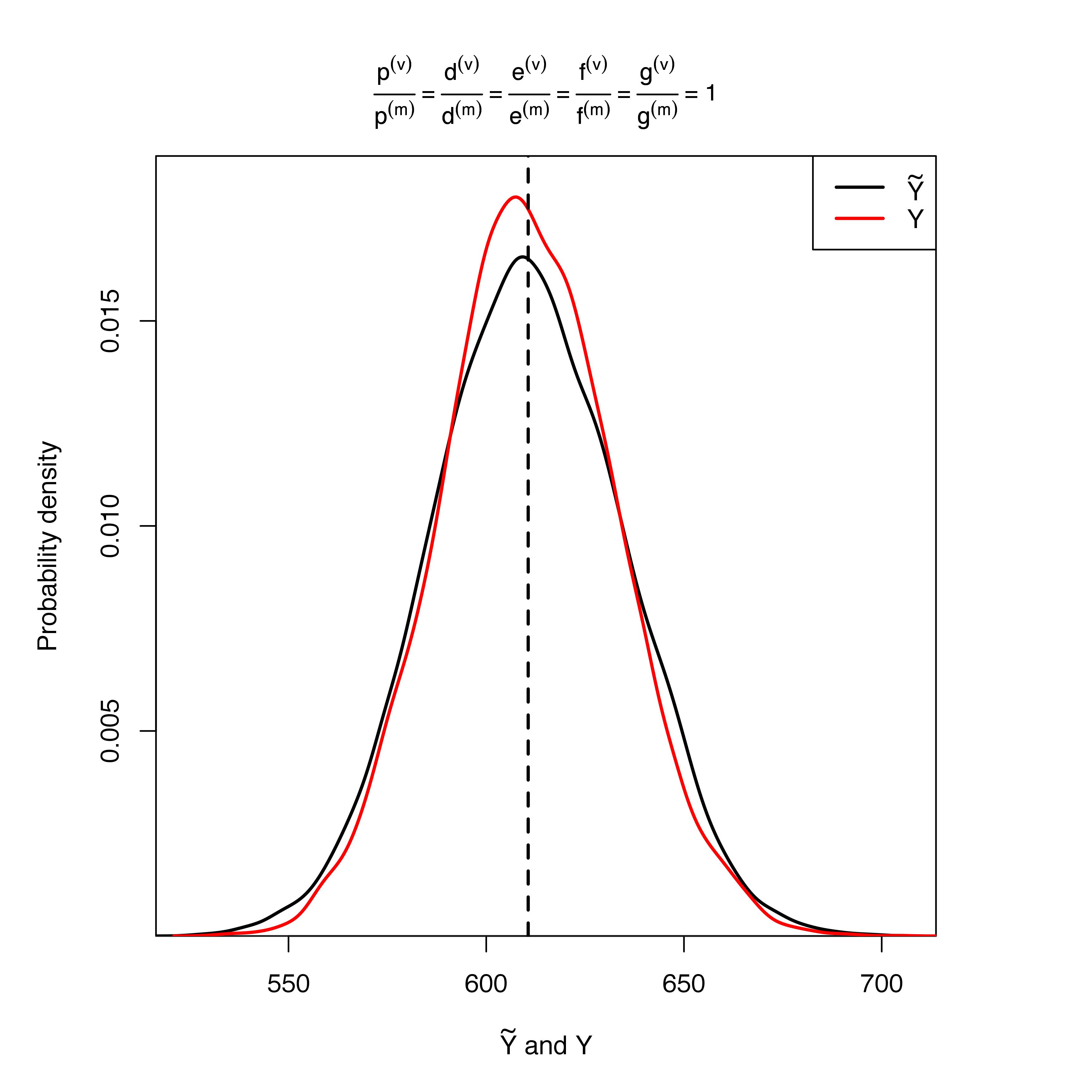}}}
\end{figure}
\begin{figure}[!ht]
\subfigure[]{\label{fig3c}}{\centerline{\includegraphics[scale=0.7]{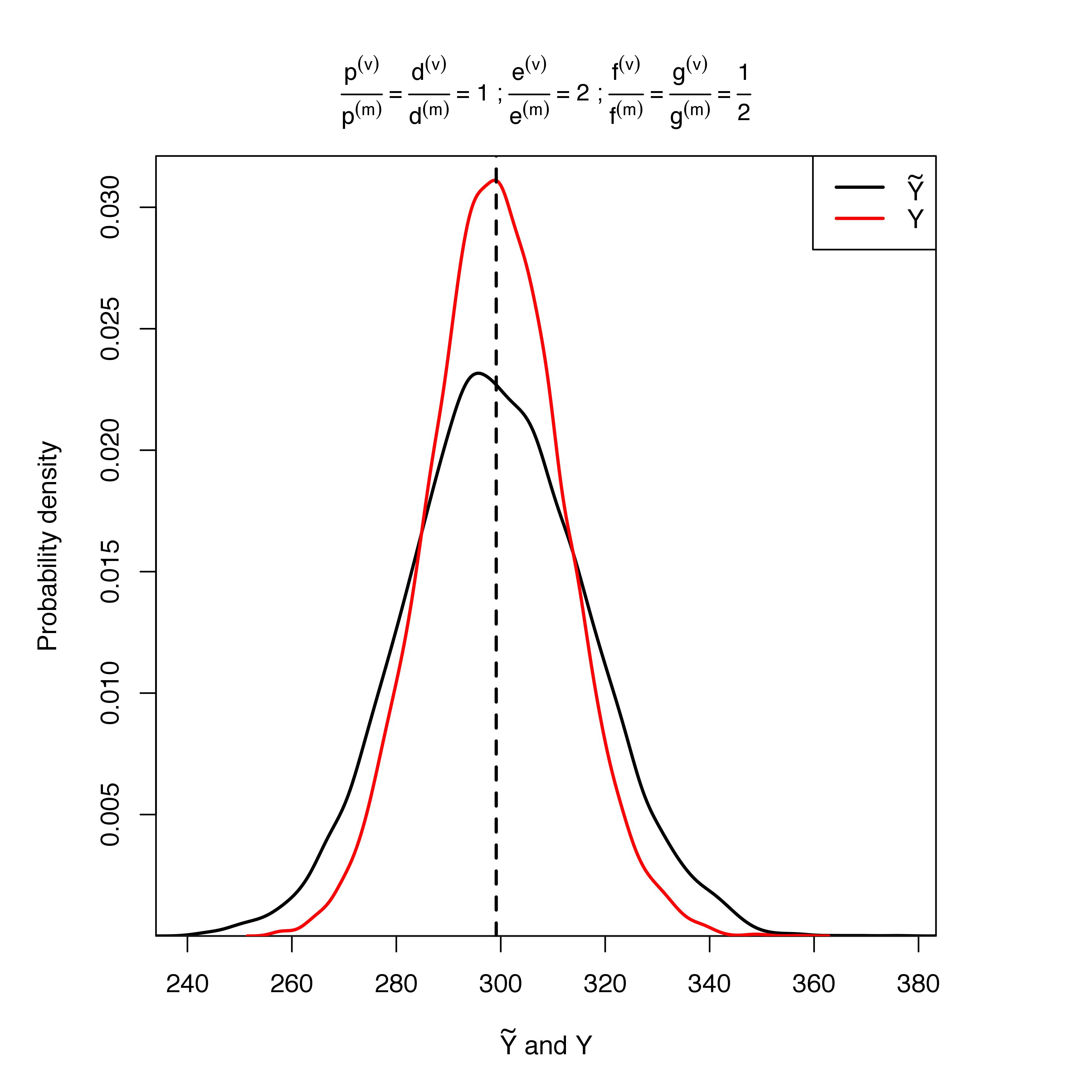}}}
\end{figure}
\begin{figure}[!ht]
\subfigure[]{\label{fig3d}}{\centerline{\includegraphics[scale=0.7]{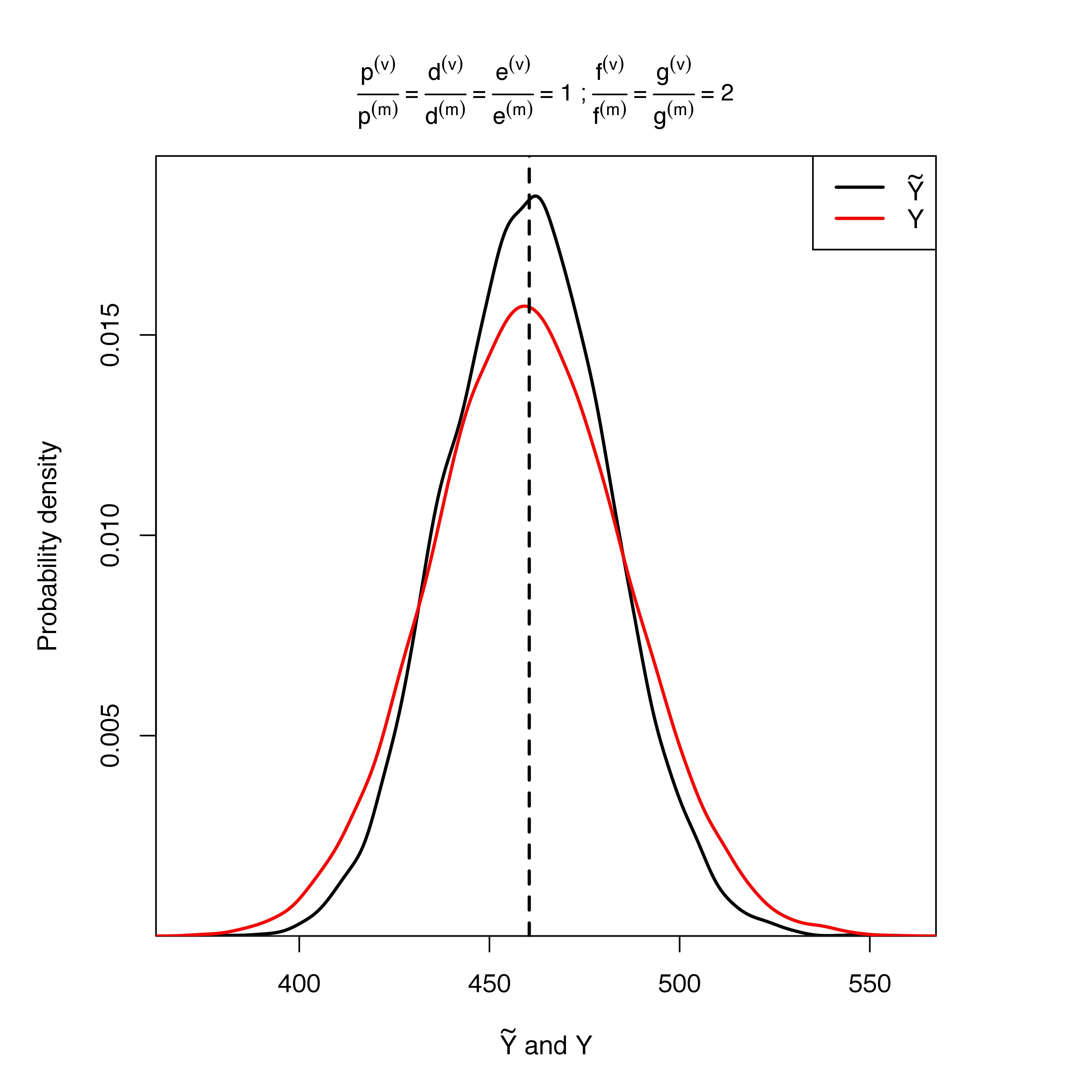}}}
\end{figure}
\begin{figure}[!ht]
\caption{\textbf{Probability density of the number of molecules at the steady state in a system without (black line) and with (red line) a buffer, in the case of protein dimerization.} The vertical dashed line represents the mean number of molecules in the steady state, which is imposed to be equal in the system with and without a buffer  ($\E(\tilde Y)=\E(Y)$).  The values of the production, degradation and interconversion parameters and the details of the numeric simulations are given in the legend of Figure 2 (a-d). (a) Unperturbed birth-death process with no degradation in the buffer; (b) Unperturbed birth-death process with degradation in the buffer; (c) Perturbed birth-death process with  little noisy interconversion gates; (d) Perturbed birth-death process with a noisy buffer. }\label{fig3}
\end{figure}

\begin{figure}[!ht]
\centerline{\includegraphics[scale=0.7]{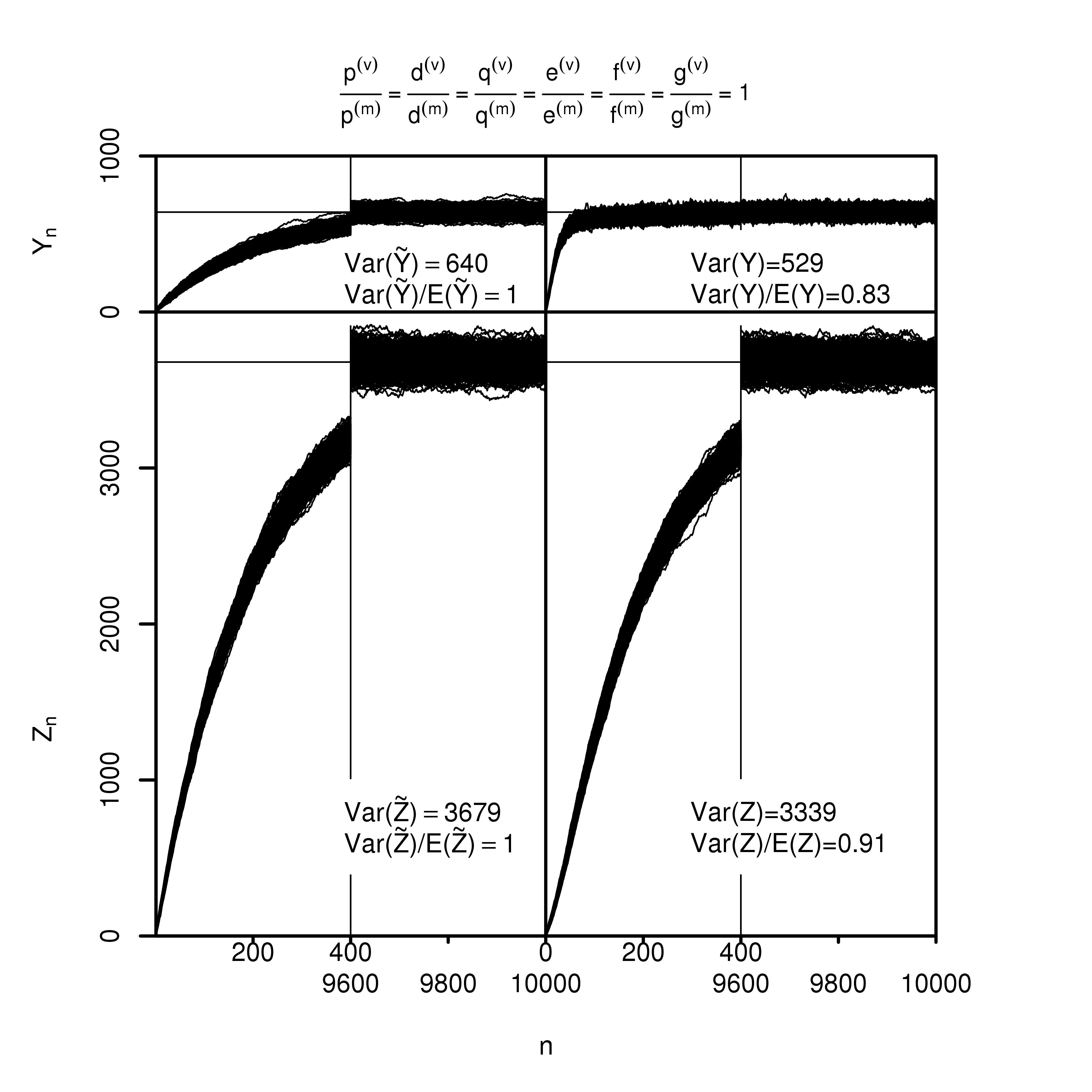}}
\caption{\textbf{Numeric simulation of the number of molecules in two systems (above and below) that are unconnected (left) and connected (right), as a function of time ($n$).} The interconversion terms are the same as for protein dimerization (Eq.(\ref{dim})). The horizontal lines represent the mean number of molecules in the steady state, which is imposed to be equal in the subsystems whether they are connected or not  ($\E(\tilde Y)=\E(Y)$ and $\E(\tilde Z)=\E(Z)$).  The parameters correspond to a usual birth-death process:  $p^{(m)}=200=p^{(v)}$, $d^{(m)}=0.05=d^{(v)}$, $q^{(m)}=100=q^{(v)}$, $e^{(m)}=0.05=e^{(v)}$, $f^{(m)}=0.0005=f^{(v)}$, $g^{(m)}=0.005=g^{(v)}$.
The production parameters of the systems without a buffer (${\tilde p^{(m)}}$ and ${\tilde q^{(m)}}$) have been determined so as to have an equal mean number of molecules in  one system in the presence or absence of the other system. The time step is $\Delta t = 0.1$, the initial conditions $Y_0 = 10 = Z_0$, the number of time steps $N = 10,000$ , and the number of runs $R=10,000$. Only the first and last iterations are shown ($n\le 400$ and $n\ge 9,600$), and  500 trajectories are drawn.}\label{fig4}
\end{figure}

\begin{figure}[!ht]
\subfigure[]{\label{fig5a}}{\centerline{\includegraphics[scale=0.7]{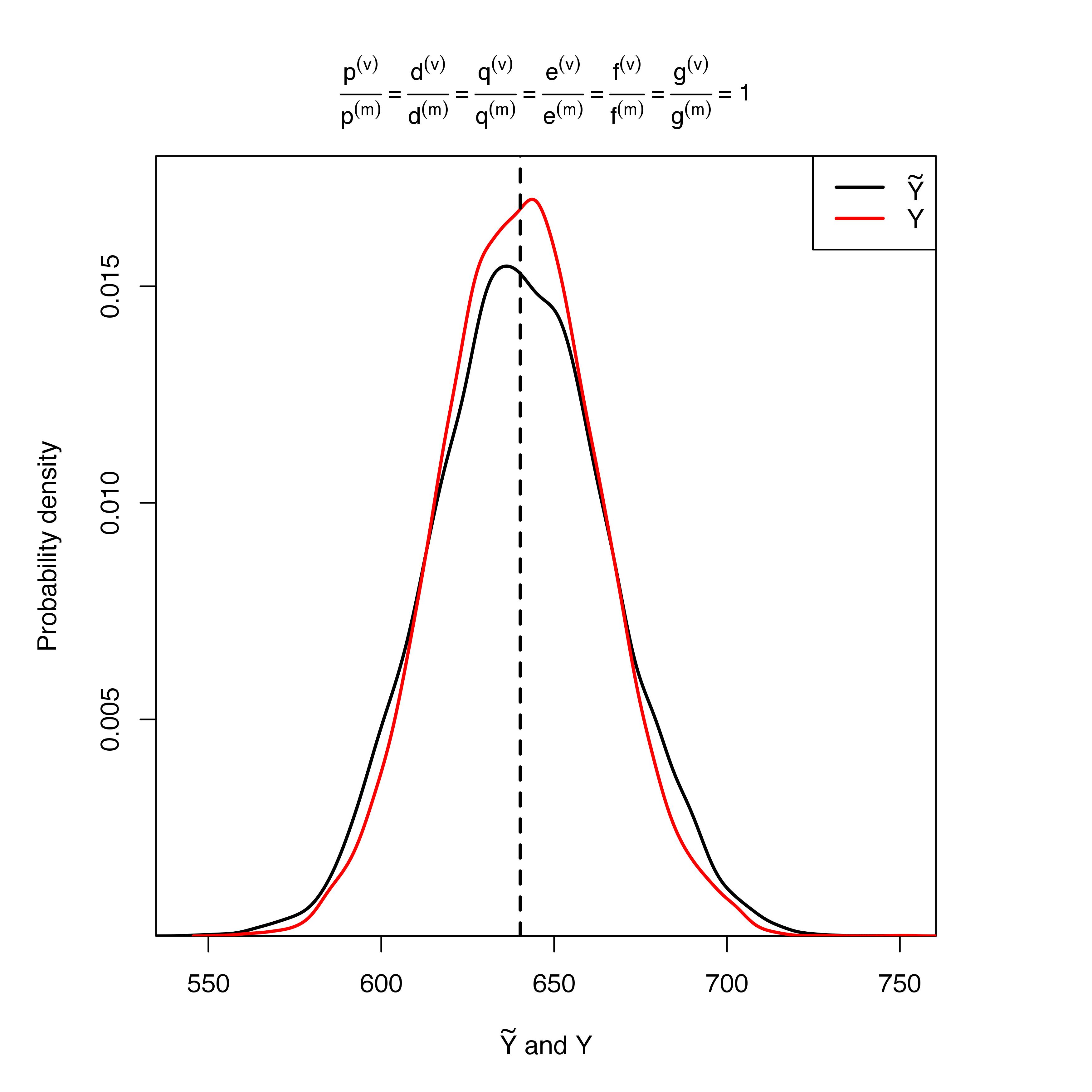}}}
\end{figure}
\begin{figure}[!ht]
\subfigure[]{\label{fig5b}}{\centerline{\includegraphics[scale=0.7]{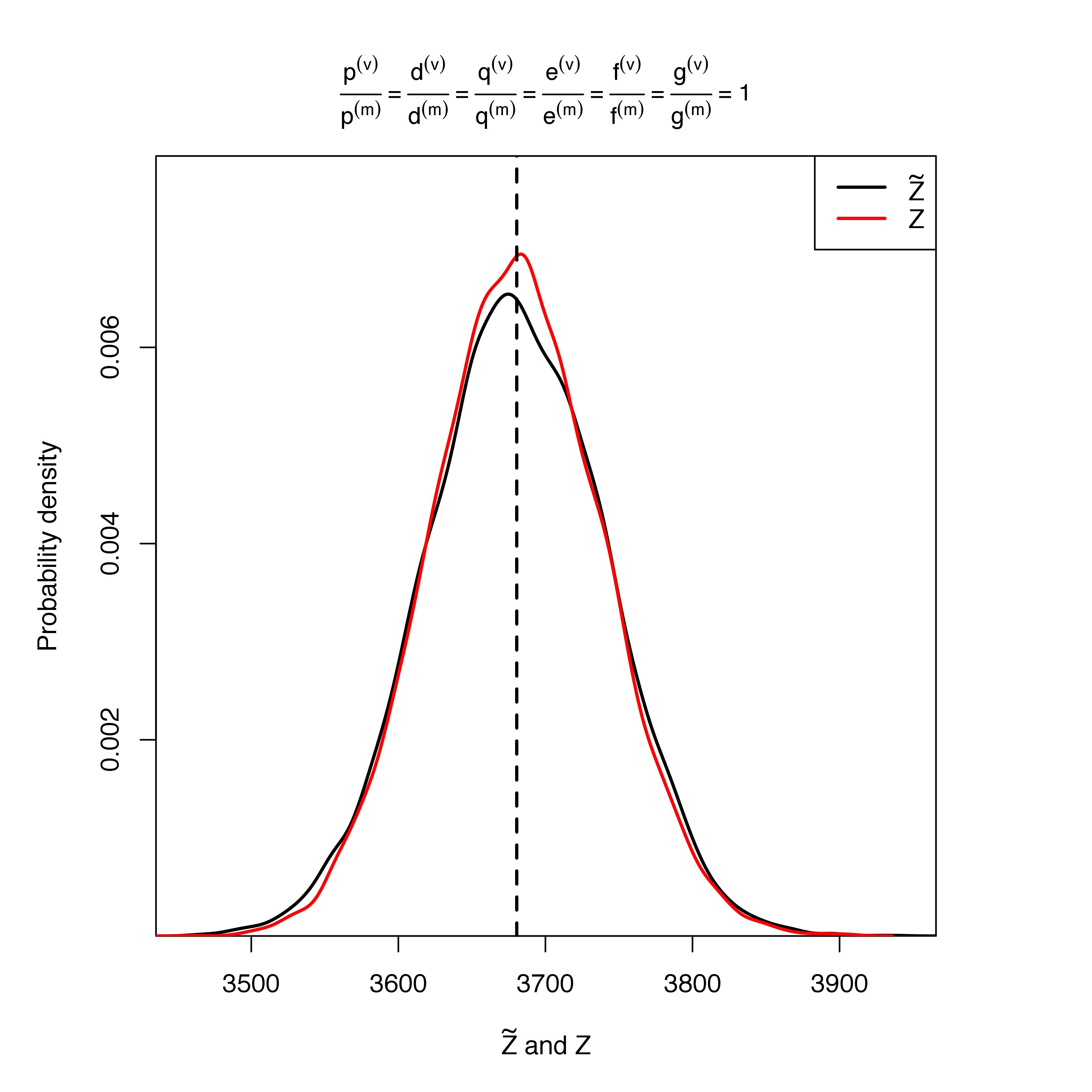}}}
\end{figure}
\begin{figure}[!ht]
\caption{\textbf{Probability density  of the number of molecules at the steady state in two systems that are unconnected  (black line) and connected (red line).} The vertical dashed line represents the mean number of molecules in the steady state, which is imposed to be equal in the subsystems whether they are connected or not  ($\E(\tilde Y)=\E(Y)$ and $\E(\tilde Z)=\E(Z)$).  The values of the production, degradation and interconversion parameters and the details of the numeric simulations are given in the legend of Figure 4. (a) First system with molecules of type $y$.  (b) Second system with molecules of type $z$.}\label{fig5}
\end{figure}

\end{document}